\documentclass[aip,amsmath,amssymb,reprint]{revtex4-2}
\usepackage[utf8]{inputenc}
\usepackage[english]{babel}

\usepackage{epsfig}
\usepackage{wrapfig}
\usepackage{bbm}
\usepackage[usenames]{color}
\usepackage{array}
\usepackage{times}
\usepackage{bm}
\usepackage[normalem]{ulem}

\usepackage{graphicx}
\usepackage{physics}
\usepackage{svg}
\usepackage{float}
\usepackage{multirow}
\usepackage[breaklinks=true]{hyperref}
\usepackage{xcolor}
\usepackage{amsmath,amsfonts,amssymb}
\usepackage{graphicx}
\usepackage{hyperref}
\usepackage{color}
\usepackage{stackengine}
\usepackage{subfigure}
\usepackage{verbatim}
\usepackage{comment}
\usepackage{stmaryrd} 

\newcommand{\divv}{\mathop{\rm div}\nolimits}

\newcommand{\eps}{\varepsilon}
\newcommand{\om}{\omega}

\renewcommand{\bm}[1]{{\bf #1}}


\newcommand{\RNum}[1]{\uppercase\expandafter{\romannumeral #1\relax}}

\begin{document}

\title{Analytical model of a Tellegen meta-atom}

\author{Daria Saltykova}
\affiliation{School of Physics and Engineering, ITMO University, Saint  Petersburg 197101, Russia}

\author{Daniel A. Bobylev}
\affiliation{School of Physics and Engineering, ITMO University, Saint  Petersburg 197101, Russia}

\author{Maxim A. Gorlach}
\email{m.gorlach@metalab.ifmo.ru}
\affiliation{School of Physics and Engineering, ITMO University, Saint  Petersburg 197101, Russia}

\begin{abstract}
Tellegen response is a nonreciprocal effect which couples electric and magnetic responses of the medium and enables unique optical properties. Here, we develop a semi-analytical model of a Tellegen particle made of magneto-optical material and explicitly compute its magnetoelectric polarizability. We demonstrate that it could substantially exceed the geometric mean of electric and magnetic polarizabilities giving rise to strong and controllable effective Tellegen response in metamaterials.
\end{abstract}

\maketitle


\textit{Tellegen medium} is a special type of material described by the constitutive relations
\begin{align} 
    &\bm{D} = \varepsilon\, \bm{E} + \chi\, \bm{H},\label{eq1}\\
    &\bm{B} = \chi\, \bm{E} + \mu\, \bm{H}.\label{eq2}
\end{align}
where real coefficient $\chi$ is termed Tellegen response or Tellegen coefficient~\cite{Tretyakov}. Equations~\eqref{eq1}-\eqref{eq2} suggest that the electric field applied to the Tellegen medium  induces magnetization, while applied magnetic field enables electric polarization. This sort of cross-coupling between electric and magnetic responses was postulated in 1948 by \mbox{B.D.H. Tellegen}~\cite{Tellegen1948} who considered a hypothetical structure where electric and magnetic dipoles were rigidly attached to each other.

Despite the conceptual simplicity of that idea, the path to experimental demonstration of Tellegen media was bumpy. As evident from Eqs.~\eqref{eq1}-\eqref{eq2}, Tellegen response requires both breaking of inversion and time-reversal symmetry. In addition, spatially homogeneous and time-independent $\chi$ is not manifested in the bulk of the medium and only modifies the boundary conditions. As a result, the very existence of Tellegen media was long debated in the  electromagnetic community.  


In condensed matter, Tellegen response was predicted~\cite{Dzyaloshinskii1960} and observed~\cite{Astrov1960} in natural media with antiferromagnetic order, e.g. Cr$_2$O$_3$, where breaking of the time-reversal symmetry occurs spontaneously. A broader class of relevant structures including  multiferroics~\cite{Pyatakov2012} and topological insulators~\cite{Armitage2016,Sekine2021} was identified later~\cite{Nenno2020}. In addition, the parallels were drawn between the equations describing Tellegen media and the electrodynamics of hypothetical  axions~\cite{Wilczek1987,Nenno2020}. However, in all of the above cases the Tellegen response is quite weak, of the order of $10^{-3}-10^{-2}$, which does not permit an efficient nonreciprocal wave manipulation.



Recently, that situation has been changed by harnessing the concept of {\it metamaterials}~-- artificial structures with engineered subwavelength periodicity. Several theoretical proposals suggested strong and controllable Tellegen response~\cite{Prudencio2023,Shaposhnikov2023,SafaeiJazi2024} and predicted that its magnitude can exceed the geometric mean of permittivity and permeability~\cite{Seidov2024}. These theoretical advances were followed by the experimental demonstration of a giant Tellegen response in the microwave range~\cite{Yang2025}, while independent theoretical~\cite{Devescovi2024} and experimental~\cite{Liu2025} studies explored the quantized Tellegen response in photonic axion insulators.

Despite these advances, a clear path to the efficient design of the Tellegen media and constituent meta-atoms is currently lacking, and the route to the maximal attainable magneto-electric coupling is practically uncharted. To clarify that and to probe the dependence of the Tellegen response on various parameters, we develop here a semi-analytical model of a Tellegen meta-atom.

\begin{figure}[b]
  \centering
\includegraphics[width=0.85\linewidth]{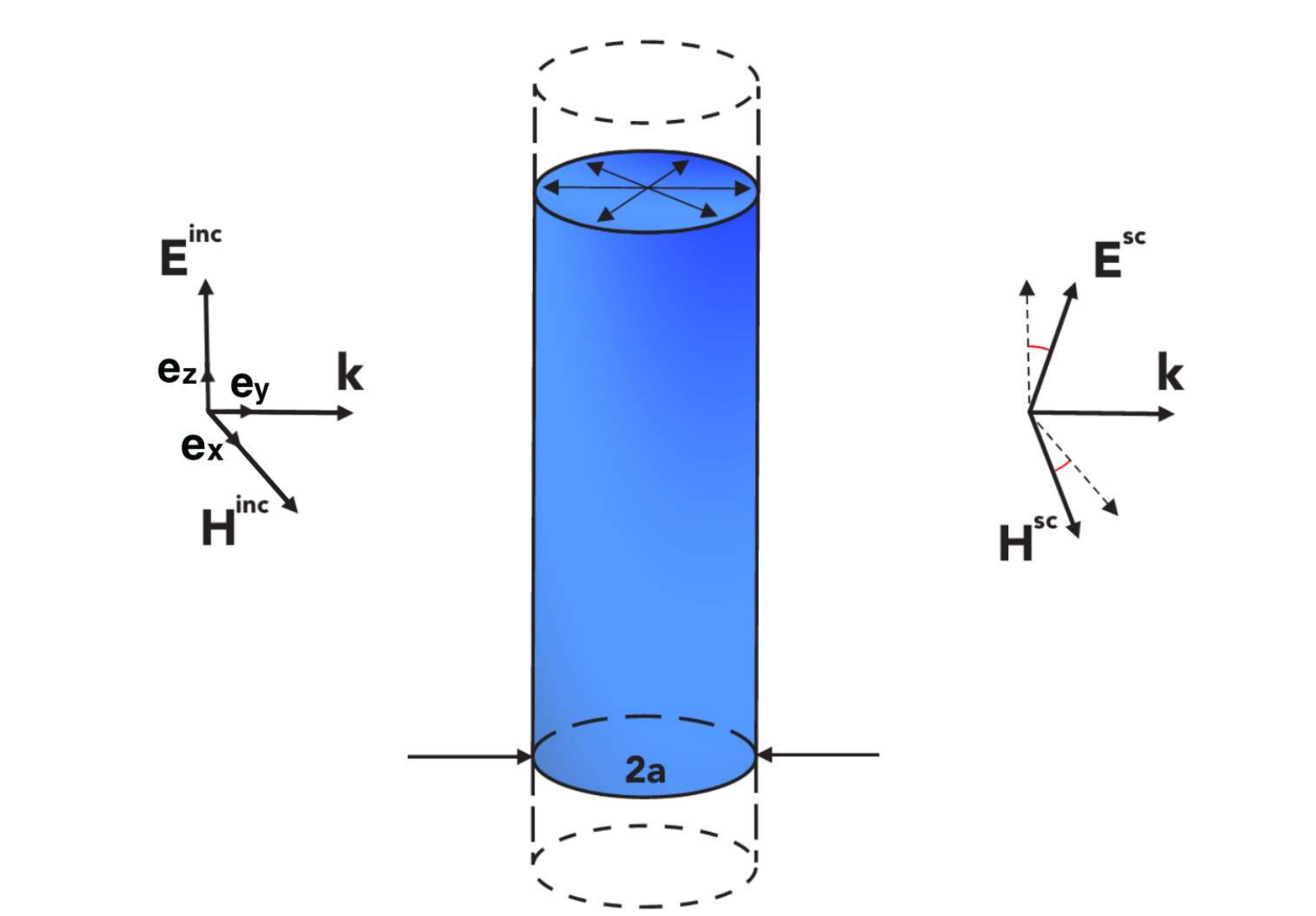}
  \caption{Scattering of a TM-polarized plane wave by an infinite radially magnetized cylinder of the radius $a$.}
  \label{schematic}
\end{figure}

Specifically, we investigate an infinite cylinder with the radial distribution of magnetization and an axis along the $Oz$ axis, Fig.~\ref{schematic}. Studying its excitation by the incident plane wave, we extract $z$-oriented electric $\bm{d}$ and magnetic $\bm{m}$ dipole moments per unit length and introduce the polarizabilities as
\begin{gather}   \bm{d}=\alpha_{ee}\,\bm{E}+\alpha_{em}\,\bm{H}\:,\label{eq:Dipole1}\\   \bm{m}=\alpha_{me}\,\bm{E}+\alpha_{mm}\,\bm{H}\:.\label{eq:Dipole2}
\end{gather}
In this setting, the Tellegen response is captured by $\alpha_{em}$ and $\alpha_{me}$ coefficients equal to each other. 

Though idealistic and difficult for the direct implementation, this system provides insights into the emergence of the effective Tellegen polarizability serving as a guide to construct more advanced systems for nonreciprocal photonics.




We start by examining the symmetry of our meta-atom. Obviously, it breaks the time-reversal symmetry $\mathcal{T}$ due to the external magnetization. At the same time, magnetization also breaks the mirror symmetry $\sigma$ in $Oxy$ plane. However, the combination of mirror symmetry and time reversal leaves the meta-atom invariant. Hence, in contrast to Ref.~\cite{SafaeiJazi2024}, the symmetry of the meta-atom cancels undesired magneto-optical effects and guarantees pure Tellegen response.

Before going into the detailed calculation, we provide a simple argument illustrating the physics behind the  Tellegen response and defining its magnitude in the long-wavelength approximation $a\ll \lambda$. We assume that the cylinder material is non-magnetic ($\mu=1$), and its  permittivity $\hat{\eps}$ is gyrotropic, i.e. the material equation inside the cylinder reads
\begin{equation} \label{constitutive relation}
    \bm{D}=\varepsilon \bm{E} +ig\bm{E} \times \bf{\hat{\rho}}
\end{equation}
where $\eps$ is the permittivity of the cylinder material, $g$ quantifies the gyrotropic response, while $\hat{\rho}$ is a unit vector in the radial direction. Note that in microwaves gyrotropic permeability $\hat{\mu}$ is more common, and the analysis and conclusions for such case are very similar.

\begin{figure}[t]
  \centering
\includegraphics[width=0.85\linewidth]{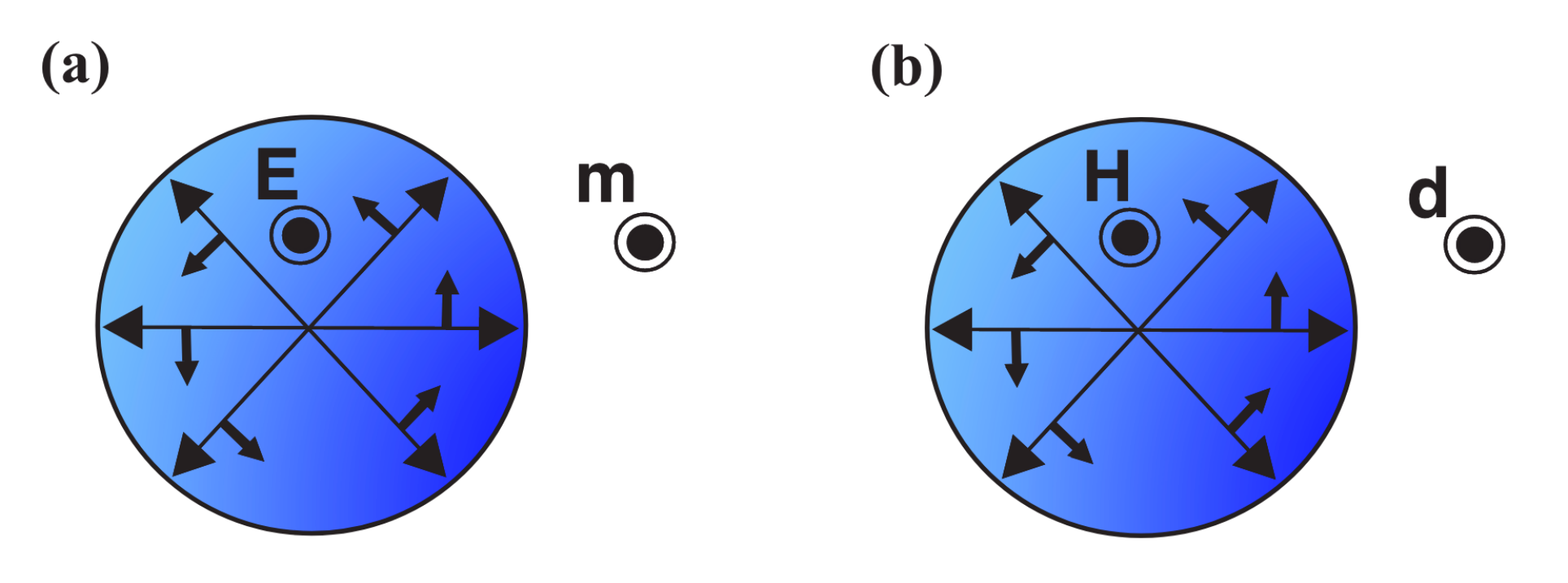}
  \caption{Qualitative reasoning of effective Tellegen response. (a) Electric field along the axis of the radially magnetized cylinder induces the vortex of polarization currents. The latter induces magnetic moment parallel to the applied electric field. (b) Oscillating magnetic field along the axis of the cylinder induces the curl of the electric field. Due to gyrotropy, this induces electric dipole moment collinear with the external magnetic field.}
  \label{Fig2}
\end{figure}

First, we assume that the external electric field ${\bf E}$ is applied along the axis of the cylinder, Fig.~\ref{Fig2}(a). In the long-wavelength approximation, electric field inside the cylinder is homogeneous and is equal to the applied field ${\bf E}$ due to the boundary conditions. Hence, electric polarizability is readily computed:
\begin{equation} \label{eq:quasi-static-e}
    \alpha_{ee}= \frac{a^2(\varepsilon-1)}{4}.
\end{equation}
In addition to that, gyrotropic permittivity Eq.~\eqref{constitutive relation} gives rise to the in-plane polarization vortex $\bm{P}=\frac{i g }{4 \pi} [\bm{E} \times \hat{\bm{\rho}}]$. This polarization oscillates in time creating a vortex of polarization currents which in turn gives rise to the magnetic moment
\begin{equation*}
\bm{m} = \frac{1}{2c} \int [\bm{r} \times \bm{j}] dV= -\frac{i q}{2} \int [\bm{\rho} \times  \bm{P}] dV= \frac{q g a^3}{12} \bm{E}\:,  
\end{equation*}
where $q=\om/c$. Thus, we recover $\alpha_{me}=\frac{q g a^3}{12}$.
%
%
Another case with applied magnetic field $\bm{H}$ along the $z$ axis is analyzed in a similar way (see Supplementary Materials, Sec.~I) resulting in $\alpha_{mm}=0$ and
\begin{gather}
\alpha_{em}=\alpha_{me}=\frac{q g a^3}{12}\:.\label{eq:quasistatic-em} 
\end{gather}

The derived magneto-electric polarizability is purely real satisfying the condition $\alpha_{em}=\alpha_{me}$, which points towards Tellegen response in our system. Furthermore, the Tellegen response Eq.~\eqref{eq:quasistatic-em} vanishes in the static limit $qa\rightarrow 0$, which motivates us to study the response of the meta-atom to the time-varying fields.


In this general case the full field $\bm{E}$ satisfies the equation
\begin{equation} \label{eq:VectorWave}
        \nabla \times \nabla \times \bm{E} = q^2 \hat{\varepsilon}\, \bm{E}\:, 
\end{equation}
where $q=\om/c$. The permittivity $\hat{\eps}$ inside the cylinder is $\phi$-dependent tensor which makes analytical solution of Eq.~\eqref{eq:VectorWave} challenging.
%

%
However, realistically, the gyrotropic correction $\propto g$ is quite small, so that the corrections $\propto g^2$ could be safely neglected. Therefore, we adopt the perturbation theory treating the gyrotropic correction to the permittivity as a perturbation and representing the electric field as
\begin{equation} \label{eq:decomposition}
        \bm{E} = \bm{E}_0+\bm{E}_1\:. 
\end{equation}
Here, $\bm{E}_0$ is the solution to the classical  scattering problem for a homogeneous dielectric cylinder with the scalar permittivity $\eps$, while $\bm{E}_1$ provides a correction to that solution proportional to the gyrotropy parameter $g$. Since $\divv\bm{E}_0=0$, an unperturbed field satisfies vector Helmhotz equation 
\begin{equation} \label{eq:VectorHelmholtz}
        \Delta \bm{E}_0 + k^2 \bm{E}_0 = 0
\end{equation}
with $k=q\,\sqrt{\eps}$. As a result, the field $\bm{E}_0$ can be expanded in the form
\begin{equation*}
    \begin{gathered}
        \bm{E}_0 = \frac{1}{\sqrt{\varepsilon}} \sum_{m=-\infty}^{+\infty} \left\{ b^{(\text{cyl})}_M(m)\bm{M}_m+ b^{(\text{cyl})}_E(m)\bm{N}_m \right\}, 
    \end{gathered}
\end{equation*}
where $b^{(\text{cyl})}_M(m)$ and $b^{(\text{cyl})}_E(m)$ are the multipole scattering coefficients further discussed in Supplementary Materials, Sec.~II.


Note that for the given TM-polarization of the incident wave (i.e. electric field along the cylinder axis),  there are no magnetic multipole coefficients: $b^{(\text{cyl})}_M(m) = 0$ for all $m$. On the contrary, in the case of TE-polarization  there are no electric multipole coefficients: $b^{(\text{cyl})}_E(m) = 0$ for all $m$.


In turn, given the multipole coefficients of the scattered field, one can readily extract the polarizabilities as
\begin{equation} \label{eq:polarizdef}
    \begin{gathered}
        \alpha_{ee}= \frac{b^{(sc)}_E(0)}{i \pi q^2 e_0},\mspace{8mu}
        \alpha_{me}= \frac{b^{(sc)}_M(0)}{i \pi q^2 e_0}\mspace{10mu}\text{(TM polarization),}\\
        \alpha_{mm}= \frac{b^{(sc)}_M(0)}{i \pi q^2 h_0}, \mspace{8mu}
        \alpha_{em}= \frac{b^{(sc)}_E(0)}{i \pi q^2 h_0}\mspace{10mu}\text{(TE polarization),}
    \end{gathered}
\end{equation}
where $e_0$ and $h_0$ are the projections of electric and magnetic fields on the axis of the cylinder. Here the scattering coefficients $ b_E(m)$, $ b_M(m)$ correspond to the unperturbed solution $\bm{E}_0$, while $ a_E(m)$ or $ a_M(m)$ are the multipole coefficients of the full field $\bm{E} = \bm{E}_0+ \bm{E}_1$.

Having an explicit solution for $\bm{E}_0$, we utilize Eq.~\eqref{eq:VectorWave} and derive the equation for the perturbation $\bm{E}_1$:
\begin{equation} \label{eq:main equation}
       \Delta\bm{E}_1 + q^2 \varepsilon \bm{E}_1 = \frac{8 \pi}{\varepsilon} \grad \rho_{\text{eff}} - \frac{4 \pi i q}{c} \bm{j}_{\text{eff}},
\end{equation}
where the effective current and charge density read
\begin{equation} \label{effcurrentcharge}
    \rho_{\text{eff}} = - \frac{1}{4 \pi} \div{(\hat{\varepsilon}_1 \bm{E}_0)}, \ \ 
    \bm{j}_{\text{eff}} = - \frac{i \omega}{4 \pi} (\hat{\varepsilon}_1  \bm{E}_0),
\end{equation}
and $\hat{\eps}_1$ is a gyrotropic correction to the permittivity.

On the other hand, the axial symmetry of the system allows to expand the field $\bm{E}_1$ in a similar form
\begin{equation}\label{eq:PerturbationExp}
    \begin{gathered}
        \bm{E}_1 = \frac{1}{\sqrt{\varepsilon}} \sum_{m=-\infty}^{+\infty} \left\{ \bm{L}_m^{(F)} + \bm{M}_m^{(G)}+\bm{N}_m^{(P)} \right\}, 
    \end{gathered}
\end{equation}
but with the different identification of the radial functions
\begin{equation}
    \begin{gathered}
        \bm{L}_m^{(F)} =  \nabla (F_m(k\rho) e^{im\phi}),\\
        \bm{M}_m^{(G)} =  \nabla \times (\hat{\bm{e}}_z G_m(k\rho) e^{im\phi}) ,\\
        \bm{N}_m^{(P)} = \frac{1}{k} \nabla \times \bm{M}_m^{(P)} =  P_m(k\rho)e^{im\phi} \hat{\bm{e}}_z  \, . 
    \end{gathered}
\end{equation}
The unknown radial functions $F_m(k\rho)$, $G_m(k\rho)$ and $P_m(k\rho)$ are regular at the coordinate origin. As we are interested in the polarizabilities Eq.~\eqref{eq:polarizdef}, we only need to examine the harmonics with $m=0$. Combining Eqs.~\eqref{eq:main equation},\eqref{eq:PerturbationExp}, we recover the equations for the radial functions
\begin{equation} \label{eq:ODE}
    \begin{gathered}
         \left(\frac{F'_0}{x} + F''_0 + F_0\right)' = 0 \, ,\\
         \left(\frac{G'_0}{x}  + G''_0 + G_0\right)' = \frac{g b^{(\text{cyl})}_E(0)}{\varepsilon}J_0(x)  \, ,\\ 
         \frac{P'_0}{x}  + P''_0 + P_0 = \frac{g b^{(\text{cyl})}_M(0)}{\varepsilon}J'_0(x) \, , 
    \end{gathered}
\end{equation}
where $x=k \rho$, $'\equiv \frac{d}{dx}$, and $J_0(x)$ is a Bessel function of the first kind which describes an unperturbed solution $\bm{E}_0$ inside the cylinder. The equation for $F_0(x)$ decouples from the rest of the system, while the multipole coefficients of the scattered field are related only to $G_0(x)$ and $P_0(x)$ functions.

Careful analysis of this problem (Supplementary Materials, Sec. III)
leads to the expressions for the polarizabilities
\begin{equation} \label{eq:polarizabilities}
    \begin{gathered}
             \alpha_{ee}=  \frac{J_0(ka)J'_0(qa) - \sqrt{\varepsilon} J_0(qa) J'_0(ka)}{i \pi q^2( \sqrt{\varepsilon} {H_0}^{(1)}(qa)J'_0(ka) -  J_0(ka) {H'_0}^{(1)}(qa))},\\
         \alpha_{mm}= \frac{J_0(ka)J'_0(qa) \sqrt{\varepsilon} -  J_0(qa) J'_0(ka)}{i \pi q^2(  {H_0}^{(1)}(qa)J'_0(ka) -  \sqrt{\varepsilon}J_0(ka) {H'_0}^{(1)}(qa))},\\
         \alpha_{me}= \frac{1}{i \pi q^2 e_0} \frac{G'_0(ka)}{\sqrt{\varepsilon} {H'_0}^{(1)}(qa)} ,\\
        \alpha_{em}= \frac{1}{i \pi q^2 h_0} \frac{P'_0(ka)}{{H'_0}^{(1)}(qa)},
    \end{gathered}
\end{equation}
where ${H_0}^{(1)}(ka)$ is the Hankel function of the first kind describing the outgoing cylindrical wave, while $e_0$ and $h_0$ are $z$-projections of the incident electric and magnetic fields for TM and TE-polarized excitation, respectively. 


In this approximation, $\alpha_{ee}$ and $\alpha_{mm}$ do not contain any contribution from the gyrotropy $g$. This is evident from the fact that these polarizabilities should keep their sign under the reversal of magnetization $g$ and hence could only include even powers of $g$.

On the contrary, magneto-electric polarizabilities $\alpha_{em}$ and $\alpha_{me}$ are linear in $g$ and capture the nonreciprocal response of our meta-atom. They include the derivatives of the respective radial functions $G_0'(ka)$ and $P_0'(ka)$ which are found numerically from Eqs.~\eqref{eq:ODE}.

\begin{figure}[b!]
  \centering
\includegraphics[width=1\linewidth]{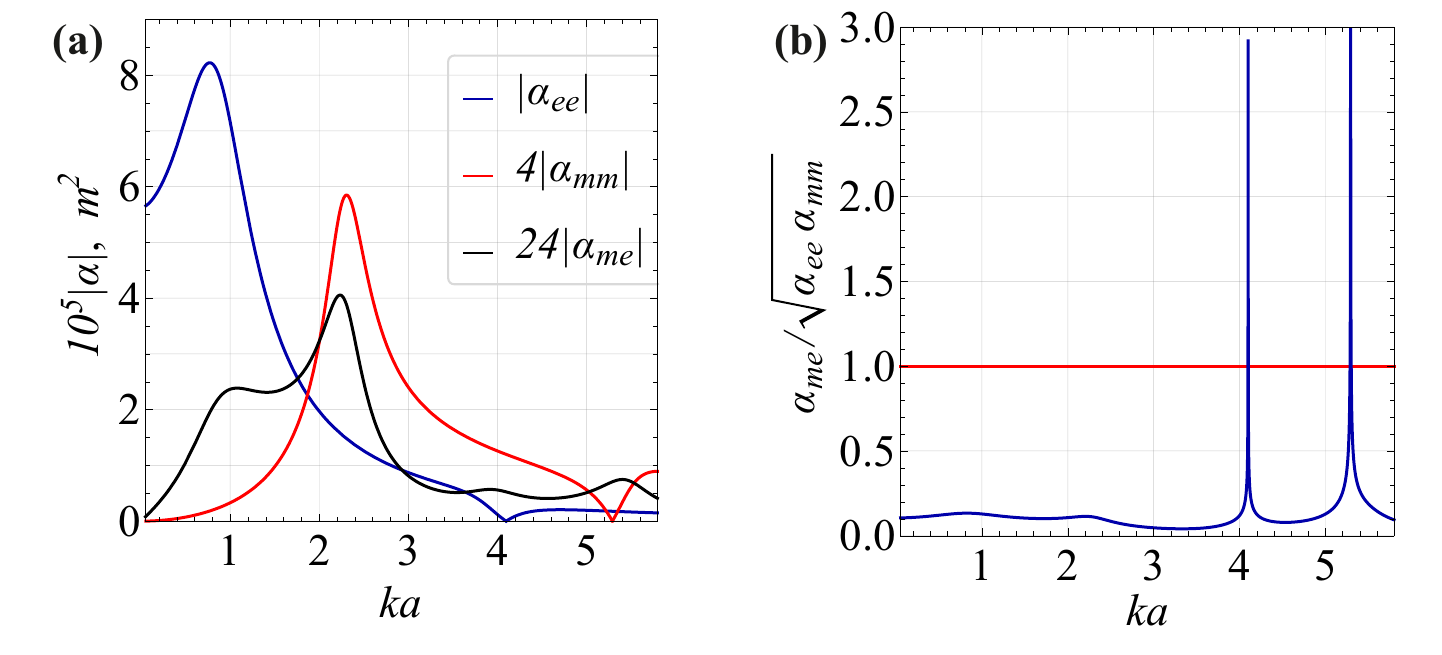} 
  \caption{Frequency dependence of polarizabilities for a radially magnetized gyrotropic cylinder.  
  (a) Electric, magnetic and magneto-electric polarizabilities versus normalized frequency $ka$. The peaks in magneto-electric coupling coincide with electric and magnetic dipole resonances. (b) Dependence of the dimensionless ratio $|\alpha^{em}| / \sqrt{|\alpha^{ee}||\alpha^{mm}|}$ on normalized frequency $ka$. At some frequencies, magneto-electric coupling exceeds previously established bound. Parameters of the system are $a=0.005$~m, $g=1$, $\varepsilon=10$, $\mu=1$.}
  \label{Fig3}
\end{figure}

Having an explicit solution for the polarizabilities, it is instructive to check their frequency dependence [Fig.~\ref{Fig3}]. It is straightforward to verify that Eqs.~\eqref{eq:polarizabilities} match Eqs.~\eqref{eq:quasi-static-e}-\eqref{eq:quasistatic-em} in the limit $qa\ll 1$, and the magneto-electric coupling vanishes in this limit (see Supplementary Materials, Sec.~IV).

Furthermore, magneto-electric polarizabilities $\alpha_{em}$ and $\alpha_{me}$ match each other at all frequencies, providing an  evidence of the Tellegen response. At non-zero frequencies, magneto-electric couplings feature both real and imaginary parts, which captures the effect of radiative losses. In turn, local maxima of magneto-electric coupling arise both at the frequencies of electric and magnetic dipole resonances manifested as peaks in electric and magnetic polarizabilities [Fig.~\ref{Fig3}(a)].

In many instances, the Tellegen polarizability is constrained by~\cite{Tretyakov}\cite{Brown1968, Shuvaev2010, Horsley2011,lindell1992methods, lindell1994electromagnetic}  
\begin{equation}\label{eq:Restriction}
|\alpha_{em}|\leq \sqrt{\alpha_{ee}\,\alpha_{mm}}\:.    
\end{equation}
Qualitatively, Eq.~\eqref{eq:Restriction} indicates that the nonzero magneto-electric coupling is necessarily  accompanied by the nonvanishing electric and magnetic responses. This bound on magneto-electric coupling was believed to be universal~\cite{Brown1968}. However, recently this limit has been challenged~\cite{Seidov2024}, while experiments demonstrated very strong Tellegen response~\cite{Yang2025}. To examine whether the bound Eq.~\eqref{eq:Restriction} breaks down for our meta-atom, in Fig.~\ref{Fig3}(b) we plot the ratio $\alpha_{em}/\sqrt{\alpha_{ee}\,\alpha_{mm}}$ versus frequency. We observe that the restriction Eq.~\eqref{eq:Restriction} is indeed violated in the two narrow frequency ranges which correspond to the vanishing electric and magnetic polarizability, while magneto-electric coupling still remains nonzero.


To independently verify the accuracy of our semi-analytical model, we simulate the scattering of TM-polarized plane wave (electric field along the $z$ axis) from a gyrotropic cylinder. To enhance the nonreciprocal effect, we assume $g=1$ in the simulation, which is typical for magnetically biased gyromagnetic materials like ferrites, but an order of magnitude larger than the gyrotropy of  magnetically biased gyroelectric materials like InSb.

The simulation is performed in the frequency domain using COMSOL Multiphysics\textsuperscript{\textregistered} software. The 2D computational domain has a radius $a+t_1+t_2$, where $a = 0.5$~cm is the cylinder radius, $t_1 = 2.25\lambda$ is the thickness of the host, $t_2 = 0.25 \lambda$ is the thickness of a perfectly matched layer (PML), and $\lambda$ is the vacuum wavelength of the incident wave. 


We examine the response of the cylinder to the TM-polarized excitation in the frequency range from $1$ to $18$~GHz with a $0.1$~GHz step. At each frequency, the cylinder produces the scattered electromagnetic field which is inspected at the points $x = \pm 2\lambda$ corresponding to the experimentally measurable forward- and back-scattered far fields. Since the incident field is polarized along the $z$ axis, the $z$ component of the scattered electric field $E_z$ is referred to as a co-polarized component, while the $z$ component of the scattered magnetic field $H_z$ captures the cross-polarized signal. In this setting, cross-polarized scattered signal provides a signature of the emergent magneto-electric coupling.

\begin{figure}[t!]
  \centering
\includegraphics[width=0.8\linewidth]{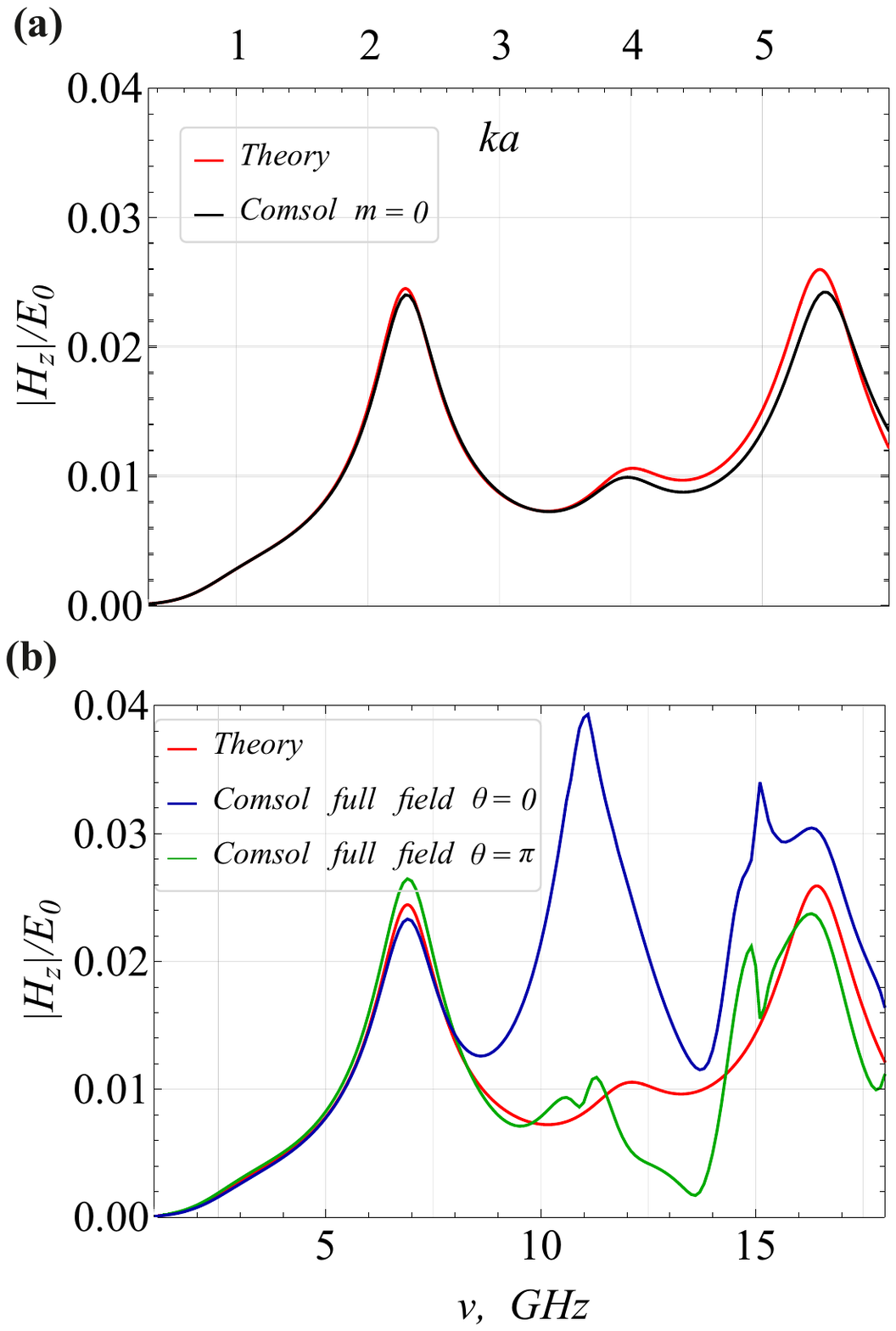}
  \caption{Full-wave numerical simulation of the scattering of a plane electromagnetic wave from a gyrotropic cylinder and its comparison with the analytical model. Parameters of the system: $a=0.005 \ \text{m}$, $g=1$, $\varepsilon=10$, $\mu=1$, the incident field is polarized along the $z$ axis, cross-polarized scattered fields $|H_z|/E_0$ are plotted at the points $x=\pm 2\lambda$. (a) Zeroth-order harmonic of the scattered field ($m=0$) versus frequency. (b) Full field scattered in forward and backward direction versus frequency.}
  \label{Fig4}
\end{figure}

First, we examine the zeroth-order ($m=0$) harmonic of the cross-polarized field $H_z$, Fig.~\ref{Fig4}(a). This component has no angular dependence $e^{i\,m\varphi}$ and thus forward and backward scattering coincide: $H_z(x=-2\lambda) = H_z(x=+2\lambda)$, so we plot the backscattering only. The retrieved frequency dependence of the cross-polarized signal is in a good agreement with the analytical model. In particular, for $ka \lesssim 3$, the relative discrepancy between the two approaches is well below 5\% which justifies the validity of our model even for the strong gyrotropy. At higher frequencies, the discrepancy increases, which is a consequence of our perturbative approach.

A richer picture is observed when we examine the contribution of the higher-order harmonics to the scattered field. Figure~\ref{Fig4}(b) illustrates that in this case forward and backward scattering no longer coincide with each other, and the zero harmonic no longer dominates the full scattered field provided the frequency $ka>3$. At the same time, the full scattered field is also different from one predicted by the analytical model. Such discrepancies are explained by the contribution from the higher-order ($m\not=0$) harmonics which is non-negligible at high frequencies. Notably, in contrast to the 3D case~\cite{SafaeiJazi2024}, the Tellegen response in a 2D geometry does not prohibit cross-polarized forward scattering as the dominant dipole moments are induced along the $z$ axis.


In summary, we have developed a semi-analytical model of a Tellegen response which arises in a cylindrical meta-atom with the radial magnetization distribution. While this particular design could be challenging to implement in practice, our analysis provides useful insights into the emergence of effective Tellegen polarizability and highlights several important aspects.

First, our model predicts that the Tellegen response arises only at non-zero frequencies, when electric and magnetic fields are intrinsically coupled to each other. Second, the maxima in the Tellegen response coincide with both electric and magnetic dipole resonances of the meta-atom. Hence, manipulating the frequencies of the respective dipole modes and making them to overlap, one could maximize the Tellegen response reaching the extreme values. Finally, as we demonstrated, the Tellegen response could be well above the standard bound $\sqrt{\alpha_{ee}\,\alpha_{mm}}$ opening a route towards giant nonreciprocity.

We believe that this study provides a useful theoretical guide for achieving strong Tellegen responses which can be further harnessed in various applications of non-reciprocal photonics.



See Supplementary Materials for the derivation of polarizabilities in the quasistatic limit, summary of the plane wave scattering on a dielectric cylinder, derivations of the polarizabilities for a gyrotropic cylinder and analysis of those expressions in the limiting case $ka\ll 1$.


Theoretical models were supported by Priority 2030 Federal Academic Leadership Program. Numerical simulations were supported by the Russian Science Foundation (Grant No.~23-72-10026).  M.A.G. acknowledges partial support from the Foundation for the Advancement of Theoretical Physics and Mathematics ``Basis''.

\section*{Author declarations}

{\bf Conflict of interest}

The authors have no conflicts to disclose.

\section*{DATA AVAILABILITY}

The data that support the findings of this study are available from the corresponding author upon reasonable request.


\bibliography{main}

\end{document}


\title{Supplementary Materials: \\ Analytical model of a Tellegen meta-atom}

\author{Daria Saltykova}
\affiliation{School of Physics and Engineering, ITMO University, Saint  Petersburg 197101, Russia}

\author{Daniel A. Bobylev}
\affiliation{School of Physics and Engineering, ITMO University, Saint  Petersburg 197101, Russia}

\author{Maxim A. Gorlach}
\affiliation{School of Physics and Engineering, ITMO University, Saint  Petersburg 197101, Russia}

\maketitle
\widetext
\tableofcontents

\section{Derivation of the quasistatic polarizabilities}

\begin{figure}[h]
  \centering
\includegraphics[width=0.4\linewidth]{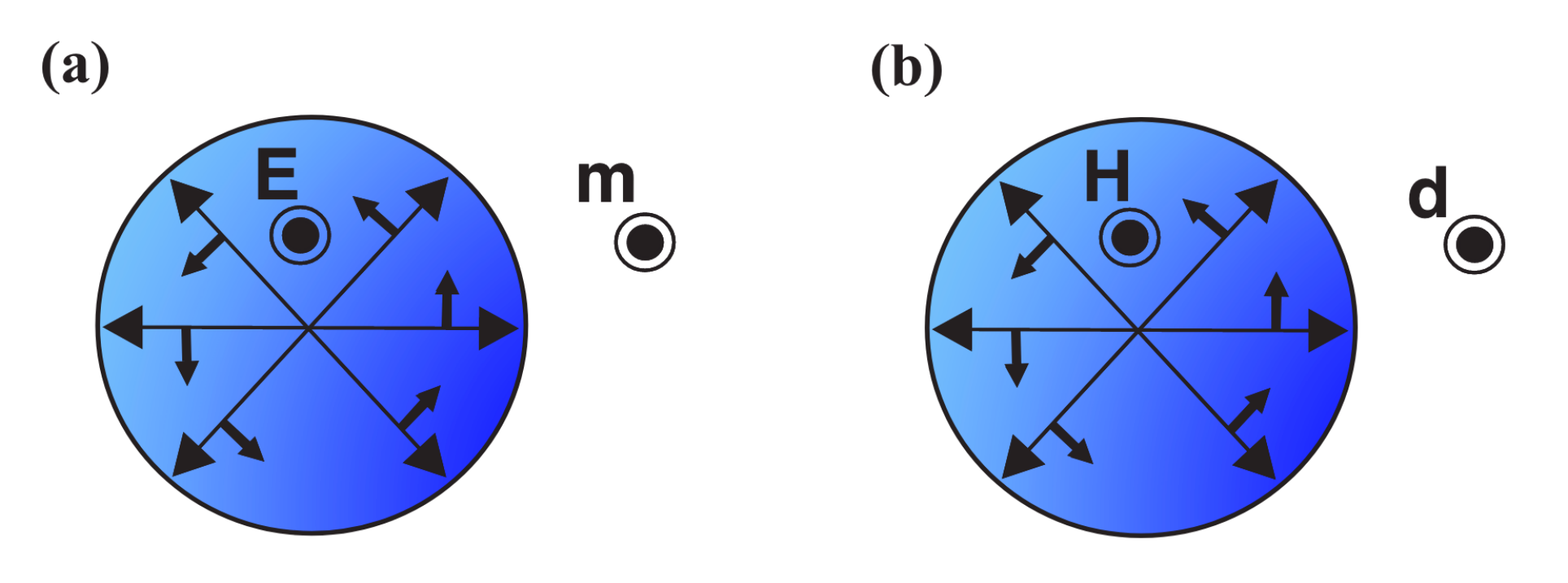}
  \caption{Qualitative reasoning of effective Tellegen response. (a) Electric field along the axis of the radially magnetized cylinder induces the vortex of polarization currents. The latter induces magnetic moment parallel to the applied electric field. (b) Oscillating magnetic field along the axis of the cylinder induces the curl of the electric field. Due to gyrotropy, this induces electric dipole moment collinear with the external magnetic field.}
  \label{Fig2}
\end{figure}

Following the logic of the main text, here we derive the quasistatic limit of polarizabilities assuming that the external magnetic field is applied along the axis of the cylinder, Fig.~\ref{Fig2}(b).

In the quasistatic limit, boundary conditions ensure that the magnetic field $\bm{H}$ is homogeneous inside the cylinder. Hence, the magnetization $\bm{M}=(\mu-1)\,\bm{H}/(4\pi)$, and the total magnetic moment per unit length is $\bm{m}=(\mu-1)\,\bm{H}\,a^2/4$. This results in magnetic polarizability
\begin{equation}
    \alpha_{mm}=\frac{a^2\,(\mu-1)}{4}\:.
\end{equation}

If the cylinder material is non-magnetic ($\mu=1$), $\alpha_{mm}$ vanishes. Next we compute magneto-electric polarizability $\alpha_{em}$. Due to the Faraday's law, time-varying magnetic field along the $z$ axis induces electric field $E_\phi$ such that $2\pi\rho\,E_\phi=i\,q\,H\,\pi\,\rho^2$, i.e.
\begin{equation}
    E_\phi=\frac{i q}{2} H \rho\:.
\end{equation}
Since the material of the cylinder is gyrotropic, $E_\phi$ component  of electric field gives rise to electric polarization along the $z$ axis:
\begin{equation}
    P_z=\frac{i g}{4 \pi} E_\phi [\bm{\hat{\phi}} \times \bm{\hat{\rho}}]_{z}= -\frac{i g}{4 \pi} E_\phi = \frac{g}{4 \pi} \frac{ q}{2} H \rho.
\end{equation}
This polarization, in turn, enables electric dipole moment
\begin{equation}
    d=\int P_z df = \frac{ g q}{8 \pi} H \int^a_0 \rho \cdot 2 \pi \rho d \rho =  \frac{ g q}{4} H \frac{a^3}{3}. 
\end{equation}
Hence, 
\begin{equation}
    d= \frac{ g q a^3}{12} H ,
\end{equation}
while the magneto-electric polarizability reads:
\begin{equation}
    \alpha_{em} = \frac{ g q a^3}{12} . 
\end{equation}
Comparing this result with the polarizability $\alpha_{me}$ in the main text, we recover $\alpha_{em} = \alpha_{me}$, as expected for the Tellegen response.



%
%

\section{Scattering of a plane wave on a dielectric cylinder}

In this section, we summarize the solution of a well-celebrated problem of the plane wave scattering on a dielectric cylinder. We expand the fields in vector cylindrical harmonics and introduce the notations further used in the main text to analyze a more complicated scattering scenario.

We adopt the CGS system of units and $e^{-i\,\omega t}$ time convention for the fields. In this case, source-free Maxwell's equations in vacuum read
\begin{equation} \label{eq:Maxwell-vacuum}
    \begin{gathered}
        \nabla \times \mathbf{E} = i\,q\,\mathbf{H}\,,\\
        \nabla \times \mathbf{H} = -i\,q\,\,\mathbf{E}\,,
    \end{gathered}
\end{equation}
where $q = \om/c$. The system~\eqref{eq:Maxwell-vacuum} is equivalent to the following one:
\begin{equation} \label{eq:curlcurl}
    \begin{gathered}
        \nabla \times \nabla \times \mathbf{E} - q^2\,\mathbf{E} = 0\,,\\
        \mathbf{H} = -\frac{i}{q}\,\nabla \times \mathbf{E}\,.\\
    \end{gathered}
\end{equation}
Since $\nabla \times \nabla \times \mathbf{E} = \nabla ( \nabla \cdot \mathbf{E} ) - \Delta \mathbf{E}$, the system~\eqref{eq:curlcurl} can be transformed accordingly:
\begin{equation} \label{eq:vector-Helmholtz}
    \begin{gathered}
        \Delta \mathbf{E} + q^2\,\mathbf{E} = 0\,,\\
        \nabla \cdot \mathbf{E} = 0\,,\\
        \mathbf{H} = -\frac{i}{q}\,\nabla \times \mathbf{E}\,,\\
    \end{gathered}
\end{equation}
where $q = \omega/c$. The system~\eqref{eq:vector-Helmholtz} contains the equation $\nabla \cdot \mathbf{E} = 0$ since it is not fulfilled automatically anymore. The first equation in the system~\eqref{eq:vector-Helmholtz}, which is the vector Helmholtz equation, can be solved using the decomposition in vector cylindrical harmonics. To construct vector harmonics, we use the following fact: if $\psi(\mathbf{r})$ is the solution to the scalar Helmholtz equation,
\begin{equation} \label{eq:scalar-Helmholtz}
    \Delta \psi + q^2\,\psi = 0\,,
\end{equation} 
the functions
\begin{equation} \label{eq:VH}
    \mathbf{L} = \nabla \psi\,, \quad \mathbf{M} =\,\nabla \times (\hat{\mathbf{a}} \psi)\,, \quad \mathbf{N} = \frac{1}{q}\,\nabla \times \mathbf{M}\,,
\end{equation}
where $\hat{\mathbf{a}}$ is a constant unit vector, are the solutions to the vector Helmholtz equation. The vector harmonics have the following properties:
\begin{equation}
    \begin{gathered}
        \nabla \times \mathbf{L} = 0\, , \quad \nabla \cdot \mathbf{M} = 0\, , \quad \nabla \cdot \mathbf{N} = 0\,.
    \end{gathered}
\end{equation}

In cylindrical coordinates, the solution to the scalar Helmholtz equation is presented in the form $\psi(\rho,\phi,z) = A F_m(q \rho) e^{i\,m\phi} e^{i\,k_z z}$, where the radial function $F_m(q \rho)$ satisfies the Bessel equation: 
\begin{equation} \label{eq:Bessel}
    \rho^2 \od[2]{F_m(q \rho)}{\rho} + \rho \od{F_m(q \rho)}{\rho} + (q^2\rho^2 - m^2) F_m(q \rho) = 0\,.
\end{equation}
We study the case when the wave vector of the incident wave is orthogonal to the axis of the cylinder. In this case, $k_z=0$ and the explicit expressions for the vector cylindrical harmonics ($\hat{\mathbf{a}} = \hat{\mathbf{z}}$) are
\begin{equation} \label{eq:VCH}
    \begin{gathered}
        \mathbf{L}_m = \nabla (F_m(q \rho) e^{im\phi}) =\left[ \od{F_m(q \rho)}{\rho} \hat{\bm{\rho}} + \frac{im}{\rho} F_m(q \rho) \hat{\bm{\phi}} \right] e^{im\phi} \, ,\\
        \mathbf{M}_m = \nabla \times (\hat{\mathbf{z}} F_m(q \rho) e^{im\phi}) = \,\left[ \frac{im}{\rho} F_m(q \rho) \hat{\bm{\rho}} - \od{F_m(q \rho)}{\rho} \hat{\bm{\phi}} \right] e^{im\phi} \, ,\\
        \mathbf{N}_m = \frac{1}{q}\,\nabla \times \mathbf{M}_m = \frac{1}{q}\,\left[ \frac{m^2}{\rho^2} F_m(q \rho) - \frac{1}{\rho} \od{F_m(q \rho)}{\rho} - \od[2]{F_m(q \rho)}{\rho}  \right] e^{im\phi} \hat{\mathbf{z}}  \, . 
    \end{gathered}
\end{equation}
The functions Eq.~\eqref{eq:VCH}, which are the vector solutions to the Helmholtz equation, are defined up to the arbitrary constant factor since the vector Helmoltz equation is linear. For convenience, we redefine vector harmonics as:
\begin{equation} \label{eq:VCH2}
    \begin{gathered}
        \tilde{\mathbf{L}}_m = \frac{1}{q}\,\left[ \od{F_m(q \rho)}{\rho} \hat{\bm{\rho}} + \frac{im}{\rho} F_m(q \rho) \hat{\bm{\phi}} \right] e^{im\phi} \, ,\\
        \tilde{\mathbf{M}}_m = \frac{i}{q} \mathbf{M}_m = \frac{i}{q}\,\left[ \frac{im}{\rho} F_m(q \rho) \hat{\bm{\rho}} - \od{F_m(q \rho)}{\rho} \hat{\bm{\phi}} \right] e^{im\phi} \, ,\\
        \tilde{\mathbf{N}}_m = \frac{1}{q} N_m =  \frac{1}{q^2}\,\left[ \frac{m^2}{\rho^2} F_m(q \rho) - \frac{1}{\rho} \od{F_m(q \rho)}{\rho} - \od[2]{F_m(q \rho)}{\rho}  \right] e^{im\phi} \hat{\mathbf{z}}\equiv F_m(q \rho)\,e^{im\phi} \hat{\mathbf{z}}  \, . 
    \end{gathered}
\end{equation}
Next, we omit tilde in these expressions. Note that in the general case different vector cylindrical harmonics can include different radial functions, each satisfying the scalar Helmholtz equation,
\begin{equation}
    \begin{gathered}
        \mathbf{L}_m^{(Q)} = \left[ \od{Q_m(q \rho)}{(q\rho)} \hat{\bm{\rho}} + \frac{im}{q\rho} Q_m(q \rho) \hat{\bm{\phi}} \right] e^{im\phi} \, ,\\
        \bm{M}_m^{(G)} = i\left[ \frac{im}{q\rho} G_m(q \rho) \hat{\bm{\rho}} - \od{G_m(q \rho)}{(q\rho)} \hat{\bm{\phi}} \right] e^{im\phi} \, ,\\
        \mathbf{N}_m^{(F)} = F_m(q \rho)\,e^{im\phi} \hat{\mathbf{z}}  \, . 
    \end{gathered}
\end{equation}
Notably, vector harmonics $\mathbf{M}_m$ and $\mathbf{N}_m$ are related to each other via $\nabla \times \mathbf{N}_m^{(G)} = - i q\,\mathbf{M}_m^{(G)}$ and $\nabla \times \mathbf{M}_m^{(F)} = i q\,\mathbf{N}_m^{(F)}$

The vector cylindrical harmonics constitute a complete set of functions and therefore the electric and magnetic fields can be expressed as their superpositions. However, in vacuum $\divv\mathbf{H}=\divv\mathbf{E}=0$, while $\divv\mathbf{L}_m\not=0$. Therefore, the expansions of electric and magnetic fields only contain $\mathbf{M}$ and $\mathbf{N}$ harmonics:
\begin{equation} \label{eq:vacuum-expansion}
    \begin{gathered}
        \mathbf{E} = \sum_{m=-\infty}^{\infty} \left\{ b_M^{(1)}(m) \mathbf{M}_m^{(1)} + b_E^{(1)}(m) \mathbf{N}_m^{(1)} \right\}\,,\\
        \mathbf{H} =  \sum_{m=-\infty}^{\infty} \left\{-b_E^{(1)}(m) \mathbf{M}^{(1)}_m + b_M^{(1)}(m) \mathbf{N}_m^{(1)}\right\}\,.
    \end{gathered}
\end{equation}
%
where the vector cylindrical harmonics with the superscript $(1)$ correspond to the volume outside the cylinder, $\rho\geq a$, while $b_{E,M}^{(1)}(m)$ denote the multipole coefficients for electric and magnetic multipoles for $\rho\geq a$.

A similar expansion is obtained in the isotropic medium with the permittivity $\varepsilon$ and permeability $\mu$:
\begin{equation} \label{eq:medium-expansions}
    \begin{gathered}
        \mathbf{E} =  \frac{1}{\sqrt{\varepsilon}} \sum_{m=-\infty}^{\infty} \left\{ b_M^{(2)}(m)\, \mathbf{M}_m^{(2)}  + b_E^{(2)}(m)\, \mathbf{N}_m^{(2)} \right\}\,,\\
         \mathbf{H} = \frac{1}{\sqrt{\mu}} \sum_{m=-\infty}^{\infty} \left\{-b_E^{(2)}(m)\, \mathbf{M}_m^{(2)} + b_M^{(2)}(m)\, \mathbf{N}_m^{(2)}\right\}\,.
    \end{gathered}
\end{equation}
%
The argument of the radial functions now contains the product $k\rho$, where  $k = q\sqrt{\eps \mu}$. As previously, superscript $(2)$ labels vector cylindrical harmonics defined inside the cylinder,  $\rho\leq a$, while $b_{E,M}^{(2)}(m)$ denote the multipole coefficients for electric and magnetic multipoles inside the cylinder.

The obtained expansions~\eqref{eq:vacuum-expansion},\eqref{eq:medium-expansions} can be used to solve a problem of scattering of a monochromatic plane wave with the frequency $\omega$ by an infinite dielectric cylinder of radius $a$,  permittivity $\eps$ and permeability $\mu$ placed in vacuum. We present the incident field in the form $\mathbf{E}^{(\mathrm{inc})} = e_0\,e^{i\,q y}\,\hat{\mathbf{z}}$. Using the identity
\begin{equation}
    e^{i\,q y} = e^{i\,q \rho \sin\phi} = \sum_{m=-\infty}^{+\infty} J_m(q \rho) e^{im\phi}\,,
\end{equation}
it is straightforward to expand the incident field into vector cylindrical harmonics:
\begin{equation}
    \begin{gathered}
        \mathbf{E}^{(\mathrm{inc})} = e_0 \sum_{m=-\infty}^{+\infty} \left\{ J_m(q \rho) e^{i\,m \phi} \hat{\mathbf{z}} \right\} \, ,\\
        \mathbf{H}^{(\mathrm{inc})} = -e_0 \sum_{m=-\infty}^{+\infty} \left\{ \frac{i}{q} \left[ \frac{i\,m}{\rho} J_m(q \rho) e^{i\,m \phi}\,\hat{\bm{\rho}} - \od{J_m(q \rho)}{\rho} e^{i\,m \phi} \hat{\bm{\phi}} \right] \right\}\,.\\
    \end{gathered}
\end{equation}
The fields scattered by the cylinder are presented as a superposition of outgoing cylindrical waves
\begin{equation} \label{scatt}
    \begin{gathered}
        \mathbf{E}^{(\mathrm{sc})} = \sum_{m=-\infty}^{+\infty} \left\{ b_M^{(\mathrm{sc})}(m) \frac{i}{q} \nabla \times[H_m^{(1)}(q \rho) e^{i\,m \phi} \hat{\mathbf{z}}] +  b_E^{(\mathrm{sc})}(m) H_m^{(1)}(q \rho) e^{i\,m \phi} \hat{\mathbf{z}} \right\} = \\ \sum_{m=-\infty}^{+\infty} \left\{ b_M^{(\mathrm{sc})}(m) \frac{i}{q} \frac{i\,m}{\rho} H_m^{(1)}(q \rho) e^{i\,m \phi} \hat{\bm{\rho}} -  b_M^{(\mathrm{sc})}(m) \frac{i}{q} \od{H_m^{(1)}(q \rho)}{\rho} e^{i\,m \phi} \hat{\bm{\phi}} +  b_E^{(\mathrm{sc})}(m) H_m^{(1)}(q \rho) e^{i\,m \phi} \hat{\mathbf{z}} \right\} \, ,\\ 
        \mathbf{H}^{(\mathrm{sc})} =  \sum_{m=-\infty}^{+\infty} \left\{- \frac{i}{q}  b_E^{(\mathrm{sc})}(m)  \nabla \times[H_m^{(1)}(q \rho) e^{i\,m \phi} \hat{\mathbf{z}}]   + b_M^{(\mathrm{sc})}(m) H_m^{(1)}(q \rho) e^{i\,m \phi} \hat{\mathbf{z}} \right\}\ = \\ =\sum_{m=-\infty}^{+\infty} \left\{- b_E^{(\mathrm{sc})}(m) \frac{i}{q}\frac{i\,m}{\rho} H_m^{(1)}(q \rho) e^{i\,m \phi} \hat{\bm{\rho}} +  b_E^{(\mathrm{sc})}(m) \frac{i}{q} \od{H_m^{(1)}(q \rho)}{\rho} e^{i\,m \phi} \hat{\bm{\phi}}  + b_M^{(\mathrm{sc})}(m) H_m^{(1)}(q \rho) e^{i\,m \phi} \hat{\mathbf{z}} \right\}\,,
    \end{gathered}
\end{equation}
while the fields inside the cylinder are
\begin{equation} \label{inside}
    \begin{gathered}
        \mathbf{E}^{(\mathrm{ins})} =  \frac{1}{\sqrt{\varepsilon}}\sum_{m=-\infty}^{+\infty} \left\{ b_M^{(\mathrm{cyl})}(m) \frac{i}{k} \nabla \times[J_m(k \rho) e^{i\,m \phi} \hat{\mathbf{z}}] +  b_E^{(\mathrm{cyl})}(m) J_m(k \rho) e^{i\,m \phi} \hat{\mathbf{z}} \right\} =  \\\frac{1}{\sqrt{\eps}}\,\sum_{m=-\infty}^{+\infty} \left\{ b_M^{(\mathrm{cyl})}(m) \frac{i}{k} \frac{i\,m}{\rho} J_m(k \rho) e^{i\,m \phi} \hat{\bm{\rho}} -  b_M^{(\mathrm{cyl})}(m) \frac{i}{k} \od{J_m(k \rho)}{\rho} e^{i\,m \phi} \hat{\bm{\phi}} + b_E^{(\mathrm{cyl})}(m) J_m(k \rho) e^{i\,m \phi} \hat{\mathbf{z}} \right\}\,,\\ 
        \mathbf{H}^{(\mathrm{ins})} =  \frac{1}{\sqrt{\mu}}\sum_{m=-\infty}^{+\infty} \left\{- \frac{i}{k}  b_E^{(\mathrm{cyl})}(m)  \nabla \times[J_m(k \rho) e^{i\,m \phi} \hat{\mathbf{z}}]   + b_M^{(\mathrm{cyl})}(m) J_m(k \rho) e^{i\,m \phi} \hat{\mathbf{z}} \right\}\ = \\\frac{1}{\sqrt{\mu}} \sum_{m=-\infty}^{+\infty} \left\{ -b_E^{(\mathrm{cyl})}(m) \frac{i}{k} \frac{i\,m}{\rho} J_m(k \rho) e^{i\,m \phi} \hat{\bm{\rho}} + b_E^{(\mathrm{cyl})}(m) \frac{i}{k} \od{J_m(k \rho)}{\rho} e^{i\,m \phi} \hat{\bm{\phi}} + b_M^{(\mathrm{cyl})}(m) J_m(k \rho) e^{i\,m \phi} \hat{\mathbf{z}} \right\}\,.
    \end{gathered}
\end{equation}
Next we match the fields inside and outside of the cylinder using the boundary conditions
$E^{(\mathrm{ins})}_{\phi,z} = E^{(\mathrm{sc})}_{\phi,z} + E^{(\mathrm{inc})}_{\phi,z}$ and $H^{(\mathrm{ins})}_{\phi,z} = H^{(\mathrm{sc})}_{\phi,z} + H^{(\mathrm{inc})}_{\phi,z}$ yields
\begin{equation}
    \begin{gathered}
        E_{\phi}: \quad \frac{1}{\sqrt
        \eps} b_M^{(\mathrm{cyl})}(m) J'_m(k a) = b_M^{(\mathrm{sc})}(m) H_m'^{(1)}(q a)\,,\\
        E_{z}: \quad \frac{1}{\sqrt{\eps}} b_E^{(\mathrm{cyl})}(m) J_m(k a) = e_0 J_m(q a) + b_E^{(\mathrm{sc})}(m) H_m^{(1)}(q a)\,,\\
        H_{\phi}: \quad \frac{1}{\sqrt{\mu}}b_E^{(\mathrm{cyl})}(m) J'_m(k a) = e_0 J'_m(q a) + b_E^{(\mathrm{sc})}(m) H_m'^{(1)}(q a)\,,\\
        H_{z}: \quad \frac{1}{\sqrt{\mu}}b_M^{(\mathrm{cyl})}(m) J_m(k a) = b_M^{(\mathrm{sc})}(m) H_m^{(1)}(q a),
    \end{gathered}
\end{equation}
where we introduced the abbreviated notation for differentiation: $J'_m(k a) = \od{J_m(k \rho)}{(k\rho)} \Huge{|}_{\small{k\rho = ka}} $ etc.
The equations for $E_{\phi}$ and $H_{z}$ ensure that $b_M^{(\mathrm{cyl})}(m) = b_M^{(\mathrm{sc})}(m) = 0$ for all $m$, i.e. the incident wave with electric field along the $z$ axis does not induce magnetic multipoles.

The equations for $E_{z}$ and $H_{\phi}$ define $b_E^{(\mathrm{cyl})}(m)$ and $b_E^{(\mathrm{sc})}(m)$:
\begin{equation} \label{eq:scattcoef1}
    \begin{gathered}
       b^{(\text{cyl})}_E(m) =  \frac{e_0 \sqrt{\varepsilon \mu}(J_m(qa)H_m'^{(1)}(q a) -  J'_m(qa) H_m^{(1)}(q a))}{ \sqrt{\mu} J_m(ka) H_m'^{(1)}(q a) - \sqrt{\varepsilon} H_m^{(1)}(q a) J'_m(ka)},\\
         b^{(\text{sc})}_E(m) =     \frac{e_0 (\sqrt{\varepsilon}J'_m(ka) J_m(qa) - \sqrt{\mu} J'_m(qa) J_m(ka))}{ \sqrt{\mu} J_m(ka) H_m'^{(1)}(q a) - \sqrt{\varepsilon} H_m^{(1)}(q a)J'_m(ka)}.
    \end{gathered}    
\end{equation}

In a similar way we analyze another polarization of the incident wave, when magnetic field is directed along the $z$ axis. In that case,
\begin{equation}
    \begin{gathered}
        \mathbf{H}^{(\mathrm{inc})} = h_0 \sum_{m=-\infty}^{+\infty} \left\{ i^mJ_m(q \rho) e^{i\,m \phi} \hat{\mathbf{z}} \right\} \, .\\
    \end{gathered}
\end{equation}
However, the boundary conditions now yield $b_E^{(\mathrm{cyl})}(m) = b_E^{(\mathrm{sc})}(m) = 0$ for all $m$ and non-zero values for
\begin{equation} \label{eq:scattcoef2}
    \begin{gathered}
       b^{(\text{cyl})}_M(m) =  \frac{i^m h_0 \sqrt{\varepsilon \mu} (J_m(qa)H_m'^{(1)}(q a) -  J'_m(qa) H_m^{(1)}(q a))}{ \sqrt{\varepsilon} J_m(ka) H_m'^{(1)}(q a) - \sqrt{\mu} H_m^{(1)}(q a) J'_m(ka)},\\
         b^{(\text{sc})}_M(m) =     \frac{i^mh_0 (\sqrt{\mu}J'_m(ka) J_m(qa) - \sqrt{\varepsilon} J'_m(qa) J_m(ka))}{ \sqrt{\varepsilon} J_m(ka) H_m'^{(1)}(q a) - \sqrt{\mu} H_m^{(1)}(q a)J'_m(ka)}.
    \end{gathered}    
\end{equation}

Having the multipole coefficients of the scattered field, one can readily extract the polarizabilities defined as electric or magnetic dipole moment in the $z$ direction per unit length of the cylinder:
\begin{equation} \label{eq:polarizdef}
    \begin{gathered}
        \alpha_{ee}= \frac{b^{(sc)}_E(0)}{i \pi q^2 e_0},\mspace{8mu}
        \alpha_{me}= \frac{b^{(sc)}_M(0)}{i \pi q^2 e_0},\\
        \alpha_{mm}= \frac{b^{(sc)}_M(0)}{i \pi q^2 h_0}, \mspace{8mu}
        \alpha_{em}= \frac{b^{(sc)}_E(0)}{i \pi q^2 h_0},
    \end{gathered}
\end{equation}
 where $\alpha_{ee}$ and $\alpha_{me}$ are recovered from the TM polarization of the incident wave (electric field directed along the axis of the cylinder),  while $\alpha_{mm}$ and $\alpha_{em}$ are computed by studying the TE illumination. This yields the explicit expressions for the polarizabilities of the dielectric cylinder:
\begin{equation} \label{eq:polarizabilitiesdielectric}
    \begin{gathered}
         \alpha_{ee}=  \frac{\sqrt{\varepsilon} J'_0(ka) J_0(qa) - \sqrt{\mu} J'_0(qa)J_0(ka)}{i \pi q^2(  \sqrt{\mu} J_0(ka) {H'_0}^{(1)}(qa) - \sqrt{\varepsilon} {H_0}^{(1)}(qa)J'_0(ka)},\\
         \alpha_{mm}= \frac{ \sqrt{\mu}  J'_0(ka) J_0(qa)  - \sqrt{\varepsilon} J'_0(qa) J_0(ka) }{i \pi q^2(\sqrt{\varepsilon}J_0(ka) {H'_0}^{(1)}(qa)) - \sqrt{\mu}  {H_0}^{(1)}(qa)J'_0(ka)) },\\
         \alpha_{me}= 0 ,\\
        \alpha_{em}= 0.
    \end{gathered}
\end{equation}
Notably, the magneto-electric coupling vanishes as guaranteed by the symmetry of the problem. 



\section{Derivation of polarizabilities for the gyrotropic cylinder}


In this section, we provide the details on the calculation of polarizabilities of a radially magnetized gyrotropic cylinder.

In analogy to the case of dielectric cylinder, the polarizabilities of a radially magnetized cylinder are defined as
\begin{equation} \label{eq:polarizdefGyr}
    \begin{gathered}
        \alpha_{ee}= \frac{a^{(sc)}_E(0)}{i \pi q^2 e_0},\mspace{8mu}
        \alpha_{me}= \frac{a^{(sc)}_M(0)}{i \pi q^2 e_0},\\
        \alpha_{mm}= \frac{a^{(sc)}_M(0)}{i \pi q^2 h_0}, \mspace{8mu}
        \alpha_{em}= \frac{a^{(sc)}_E(0)}{i \pi q^2 h_0},
    \end{gathered}
\end{equation}
where $e_0$ and $h_0$ are the $z$-projections of electric and magnetic fields of TM and TE incident waves. TM- and TE-polarized illumination is used to extract $\alpha_{ee}$, $\alpha_{me}$ and $\alpha_{mm}$, $\alpha_{em}$, respectively.

The quantities $ a_E(m)=b_E(m)+\tilde{a}_E(m)$ and $ a_M(m) = b_M(m)+\tilde{a}_M(m)$ are the multipole coefficients of the full scattered field $\bm{E} = \bm{E}_0+ \bm{E}_1$. In turn, $b_E(m)$ and $b_M(m)$ are the multipole scattering coefficients for the dielectric cylinder, while $\tilde{a}_E(m)$ and $\tilde{a}_M(m)$ provide first-order corrections proportional to the gyrotropy $g$ of the cylinder.



\begin{figure}[t]
  \centering
\includegraphics[width=0.5\linewidth]{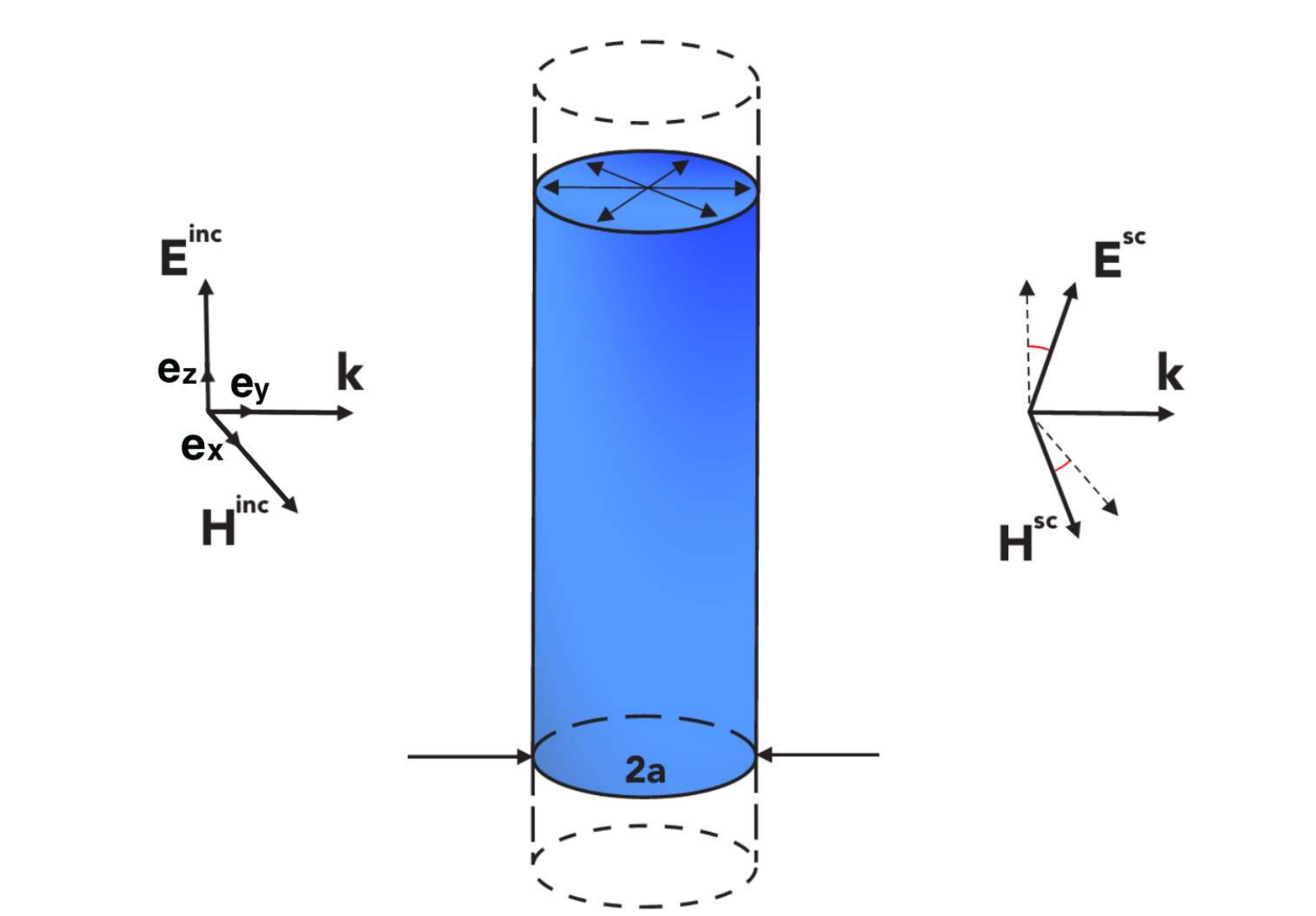}
  \caption{Scattering of a TM-polarized plane wave by an infinite radially magnetized cylinder of radius $a$.}
  \label{fig:S2}
\end{figure}

We start by examining TM polarization of the incident field (see Fig.~\ref{fig:S2}). In analogy to the problem discussed in Sec.~II, we expand the incident field into vector cylindrical harmonics:
\begin{equation}
    \begin{gathered}
        \mathbf{E}^{(\mathrm{inc})} = e_0 \sum_{m=-\infty}^{+\infty} \left\{ J_m(q \rho) e^{i\,m \phi} \hat{\mathbf{z}} \right\} \, ,\\
        \mathbf{H}^{(\mathrm{inc})} = -e_0 \sum_{m=-\infty}^{+\infty} \left\{ \frac{i}{q} \left[ \frac{i\,m}{\rho} J_m(q \rho) e^{i\,m \phi}\,\hat{\bm{\rho}} - \od{J_m(q \rho)}{\rho} e^{i\,m \phi} \hat{\bm{\phi}} \right] \right\}\,.\\
    \end{gathered}
\end{equation}
Next we present the electric field inside and outside of the cylinder in the form:
\begin{align} \label{asymptE}
    &\mathbf{E}^{(\mathrm{ins})} = \mathbf{E}_0^{(\mathrm{ins})} + \mathbf{E}_1^{(\mathrm{ins})}, \\
    & \mathbf{E}^{(\mathrm{sc})} = \mathbf{E}_0^{(\mathrm{sc})} + \mathbf{E}_1^{(\mathrm{sc})},
\end{align}
where $\mathbf{E}_0^{(\mathrm{ins/sc})}$ provides known solution for the dielectric cylinder (Eq.~\eqref{inside}, \eqref{scatt}) and $\mathbf{E}_1^{(\mathrm{ins/sc})}$ is a correction to that solution linear in gyrotropy $g$. Because of the axial symmetry of the system, this correction can also be expanded into cylindrical harmonics:
\begin{equation}\label{eq:PerturbationExpGyrotropic}
    \begin{gathered}
        \bm{E}^{(\mathrm{ins})}_1 = \frac{1}{\sqrt{\varepsilon}} \sum_{m=-\infty}^{+\infty} \left\{ \bm{L}_m^{(F)} + \bm{M}_m^{(G)}+\bm{N}_m^{(P)} \right\} =  \\\frac{1}{\sqrt{\varepsilon}} \sum_{m=-\infty}^{+\infty} \left\{ \nabla (F_m(k\rho) e^{im\phi}) +  \nabla \times (\hat{\bm{e}}_z G_m(k\rho) e^{im\phi})+ P_m(k\rho)e^{im\phi} \hat{\bm{e}}_z  \right\} , \\
         \bm{E}^{\mathrm{(sc)}}_1 = \sum_{m=-\infty}^{+\infty} \left\{ \tilde{a}^{(sc)}_L(m) \nabla (H_m^{(1)}(q \rho) e^{im\phi}) +\tilde{a}_M^{(\mathrm{sc})}(m) \frac{i}{q} \nabla \times[H_m^{(1)}(q \rho) e^{i\,m \phi} \hat{\bm{e}}_z] +  \tilde{a}_E^{(\mathrm{sc})}(m) H_m^{(1)}(q \rho) e^{i\,m \phi} \hat{\bm{e}}_z \right\}.\\
    \end{gathered}
\end{equation}

Note that the radial functions for the solution inside are modified, while the structure of the outside solution is captured by the outgoing cylindrical waves associated with the Hankel function of the first kind.

We perform a similar expansion for the magnetic fields:
\begin{align} \label{asymptH}
    &\mathbf{H}^{(\mathrm{ins})} = \mathbf{H}_0^{(\mathrm{ins})} + \mathbf{H}_1^{(\mathrm{ins})}, \\
    & \mathbf{H}^{(\mathrm{sc})} = \mathbf{H}_0^{(\mathrm{sc})} + \mathbf{H}_1^{(\mathrm{sc})}.
\end{align}
Here $\mathbf{H}_0^{(\mathrm{ins/sc})}$ is a solution for the dielectric cylinder (Eqs.~\eqref{inside}, \eqref{scatt}) and $\mathbf{H}_1^{(\mathrm{ins/sc})}$ is a first-order correction, which has the form:
\begin{equation}\label{eq:PerturbationExp}
    \begin{gathered}
       \mathbf{H}^{(\mathrm{ins})}_1 = - \frac{i}{q} \rot \mathbf{E}^{(\mathrm{ins})}_1 = - \frac{i}{q} \frac{1}{\sqrt{\varepsilon}}  \sum_{m=-\infty}^{+\infty} \left\{ \rot \bm{L}_m^{(F)} + \rot \bm{M}_m^{(G)}+\rot \bm{N}_m^{(P)} \right\} =  \\
        - \frac{i}{k}   \sum_{m=-\infty}^{+\infty} \left\{ i k \bm{N}_m^{(G)}- i k \bm{M}_m^{(P)} \right\} =      \sum_{m=-\infty}^{+\infty} \left\{ G_m(k\rho)e^{im\phi} \hat{\bm{e}}_z- \nabla \times (\hat{\bm{e}}_z P_m(k\rho) e^{im\phi})  \right\} , \\
         \mathbf{H}^{\mathrm{(sc)}}_1 = \sum_{m=-\infty}^{+\infty} \left\{- \frac{i}{q} \tilde{a}_E^{(\mathrm{sc})}(m)  \nabla \times[H_m^{(1)}(q \rho) e^{i\,m \phi} \hat{\mathbf{z}}]   + \tilde{a}_M^{(\mathrm{sc})}(m)  H_m^{(1)}(q \rho) e^{i\,m \phi} \hat{\mathbf{z}} \right\} .
    \end{gathered}
\end{equation}

The solutions inside and outside are matched to each other via the boundary conditions
$E^{(\mathrm{ins})}_{\phi,z} = E^{(\mathrm{sc})}_{\phi,z} + E^{(\mathrm{inc})}_{\phi,z}$ and $H^{(\mathrm{ins})}_{\phi,z} = H^{(\mathrm{sc})}_{\phi,z} + H^{(\mathrm{inc})}_{\phi,z}$,
which require the continuity of $\varphi$ and $z$ components of electric and magnetic field. This yields
\begin{equation}
    \begin{gathered} \label{system}
        E_{\phi}: \quad \frac{1}{\sqrt
        \eps} [b_M^{(\mathrm{cyl})}(m) J'_m(k a) + G'_m(ka) - \frac{m}{ka} F_m]= [b_M^{(\mathrm{sc})}(m)+  \tilde{a}_M^{(\mathrm{sc})}(m)]H_m'^{(1)}(q a) - \tilde{a}_L^{(\mathrm{sc})}(m)\frac{m}{ka} H_m^{(1)}(q a) \,,\\
        E_{z}: \quad \frac{1}{\sqrt{\eps}} [b_E^{(\mathrm{cyl})}(m) J_m(k a) + P_m(k a)] = e_0 J_m(q a) + [b_E^{(\mathrm{sc})}(m) + \tilde{a}_E^{(\mathrm{sc})}(m)]H_m^{(1)}(q a)\,,\\
        H_{\phi}: \quad b_E^{(\mathrm{cyl})}(m) J'_m(k a) + P'_m(k a) = e_0 J'_m(q a) + [b_E^{(\mathrm{sc})}(m)+\tilde{a}_E^{(\mathrm{sc})}(m)] H_m'^{(1)}(q a)\,,\\
        H_{z}: \quad b_M^{(\mathrm{cyl})}(m) J_m(k a) -\frac{G'_m(ka)}{ka}-G''_m(ka)+\frac{m^2}{k^2a^2}G_m(ka) = [b_M^{(\mathrm{sc})}(m) +\tilde{a}_M^{(\mathrm{sc})}(m)]H_m^{(1)}(q a).
    \end{gathered}
\end{equation}

Given the polarization of the incident field, $b_M^{(\mathrm{cyl})}(m) = b_M^{(\mathrm{sc})}(m) = 0$ for all $m$ (see Sec. II). Furthermore, to compute the polarizabilities, we only need the multipole coefficients with $m=0$ [see Eq.~\eqref{eq:polarizdefGyr}]. Taking this into account, we rewrite the system \eqref{system} in the form:
\begin{equation}
    \begin{gathered} \label{system2}
        E_{\phi}: \quad \frac{1}{\sqrt
        \eps}  G'_0(ka) =   \tilde{a}_M^{(\mathrm{sc})}(0)H_0'^{(1)}(q a) \,,\\
        E_{z}: \quad \frac{1}{\sqrt{\eps}} [b_E^{(\mathrm{cyl})}(0) J_0(k a) + P_0(k a)] = e_0 J_0(q a) + [b_E^{(\mathrm{sc})}(0) + \tilde{a}_E^{(\mathrm{sc})}(0)]H_0^{(1)}(q a)\,,\\
        H_{\phi}: \quad b_E^{(\mathrm{cyl})}(0) J'_0(k a) + P'_0(k a) = e_0 J'_0(q a) + [b_E^{(\mathrm{sc})}(0)+\tilde{a}_E^{(\mathrm{sc})}(0)] H_0'^{(1)}(q a)\,,\\
        H_{z}: \quad  -\frac{G'_0(ka)}{ka}-G''_0(ka) = \tilde{a}_M^{(\mathrm{sc})}(0)H_0^{(1)}(q a),
    \end{gathered}
\end{equation}
In particular, the first and the fourth equations of the system \eqref{system2} read:
\begin{equation}
    \begin{gathered} \label{system3}
        E_{\phi}: \quad \frac{1}{\sqrt
        \eps}  \tilde{a}_E^{(\mathrm{cyl})}(0)G'_0(ka) =   \tilde{a}_M^{(\mathrm{sc})}(0)H_0'^{(1)}(q a) \,,\\
        H_{z}: \quad  \tilde{a}_E^{(\mathrm{cyl})}(0)(-\frac{G'_0(ka)}{ka}-G''_0(ka)) = \tilde{a}_M^{(\mathrm{sc})}(0)H_0^{(1)}(q a),
    \end{gathered}
\end{equation}
Obviously, the unknown radial function $G_0(x)$ can be multiplied by the arbitrary number and redefined as $\tilde{a}_E^{(\mathrm{cyl})}(0)G_0 \rightarrow G_0$. By doing so with Eqs.~\eqref{system3}, we recover
\begin{equation} \label{a_m}
 \begin{cases}
   \tilde{a}_M^{(\mathrm{sc})}(0) = \frac{G'_0(ka)}{\sqrt {\eps} H_0'^{(1)}(q a)},\\
   \sqrt{\eps} H_0'^{(1)}(q a)( \frac{G'_0}{ka}+G''_0(ka))+H_0^{(1)}(q a)G'_0(ka)=0.
 \end{cases}
\end{equation}
The first equation in the system Eq.~\ref{a_m} defines the multipole scattering coefficient for the field outside in terms of the radial function $G_0$. The second equation follows from the requirement that the determinant of the system Eq.~ \label{system3} is zero.


On the other hand, the unknown radial functions satisfy the system of the differential equations
\begin{equation} \label{ODE}
    \begin{gathered}
         \left(\frac{F'_0}{x} + F''_0 + F_0\right)' = 0 \, ,\\
         \left(\frac{G'_0}{x}  + G''_0 + G_0\right)' = \frac{g b^{(\text{cyl})}_E(0)}{\varepsilon}J_0(x)  \, ,\\ 
         \frac{P'_0}{x}  + P''_0 + P_0 = \frac{g b^{(\text{cyl})}_M(0)}{\varepsilon}J'_0(x) \, , 
    \end{gathered}
\end{equation}
In the case of TM polarization of the incident field, this system simplifies:
\begin{equation} \label{ODE2}
    \begin{gathered}
         \left(\frac{F'_0}{x} + F''_0 + F_0\right)' = 0 \, ,\\
         \left(\frac{G'_0}{x}  + G''_0 + G_0\right)' = \frac{g b^{(\text{cyl})}_E(0)}{\varepsilon}J_0(x)  \, ,\\ 
         \frac{P'_0}{x}  + P''_0 + P_0 = 0\, , 
    \end{gathered}
\end{equation}
where we utilized the fact that $b_M^{(\mathrm{cyl})}(m) = b_M^{(\mathrm{sc})}(m) = 0$ for all $m$.
Finally, we can write the system as
\begin{equation} \label{a_mfinal}
 \begin{cases}
   \tilde{a}_M^{(\mathrm{sc})}(0) = \frac{G'_0(ka)}{\sqrt {\eps} H_0'^{(1)}(q a)},\\
   \left(\frac{G'_0}{x}  + G''_0 + G_0\right)' = \frac{g b^{(\text{cyl})}_E(0)}{\varepsilon}J_0(x), \\
    \sqrt{\eps} H_0'^{(1)}(q a)( \frac{G'_0}{ka}+G''_0(ka))+H_0^{(1)}(q a)G'_0(ka)=0.
 \end{cases}
\end{equation}
Based on the first equation in Eq.~\eqref{a_mfinal}, to find $\tilde{a}_M^{(\mathrm{sc})}(0)$, we need to calculate the function $G'_0(x)$. To do so, we solve the differential equation for the function $G_0(x)$, taking into account the boundary condition:
\begin{equation} \label{eqforG_0}
 \begin{cases}
   \left(\frac{G'_0}{x}  + G''_0 + G_0\right)' = \frac{g b^{(\text{cyl})}_E(0)}{\varepsilon}J_0(x), \\
    \sqrt{\eps} H_0'^{(1)}(q a)( \frac{G'_0}{ka}+G''_0(ka))+H_0^{(1)}(q a)G'_0(ka)=0.
 \end{cases}
\end{equation}
To proceed, we recast the equation for the radial function as 
\begin{equation}\label{eqforG2}
    \frac{G'_0}{x}  + G''_0 + G_0 = \frac{g b^{(\text{cyl})}_E(0)}{\varepsilon} \int^{x}_0 J_0(t) dt,
\end{equation}
and seek the solution in the form
\begin{equation} \label{formofG}
  G_0(x) = B\,J_0(x)+A+f(x),
\end{equation}
where $B\,J_0(x)$ is a homogeneous solution of Eq.~\ref{eqforG2}, $f(x)$ is a particular solution to the inhomogeneous problem with the zero value at $x=0$, and $A$ is an inessential constant that drops out from the expression for the multipole coefficient. The unknown $B$ constant is found from the boundary condition [see the second expression in Eq.~\eqref{eqforG_0}] and takes the form:
\begin{equation} \label{eq:polarizabilities}
    \begin{gathered}
        B= - \frac{g b^{(\text{cyl})}_E(0)}{\varepsilon}  \frac{ \sqrt{\varepsilon} {H'_0}^{(1)}(qa)  (\frac{f'(ka)}{ka}+f''(ka))+  {H_0}^{(1)}(qa) f'(ka)}{\sqrt{\varepsilon} {H'_0}^{(1)}(qa)  (\frac{J'_0(ka)}{ka}+J''_0(ka))+  {H_0}^{(1)}(qa) J'_0(ka)}.\\
    \end{gathered}
\end{equation}

The function $f(x)$ has to be calculated numerically. Due to the structure of Eq.~\eqref{eqforG2}, its numerical solution around $x=0$ can be problematic, so it is necessary to find the initial conditions at a point near zero. We therefore seek the solution for $f(x)$ in the form of series in powers of $x$:
\begin{equation} \label{functionseries}
    \begin{gathered}
       f(x) = \sum^{\infty}_{m=0}b_{2m+1}x^{2m+1}
    \end{gathered}
\end{equation}
By matching the left and right sides of the equation \ref{eqforG2}, presented as series, we can consistently find the coefficients $b_{2m+1}$. The leading-order terms read:
\begin{equation} \label{functionf}
    \begin{gathered}
       f(x) \approx \frac{x^3}{9}-\frac{7 x^5}{900}+...  \ 
    \end{gathered}
\end{equation}
With the expansion Eq.~\eqref{functionf}, we can calculate the initial condition at any point close to zero and numerically find the function $f(x)$ in the entire range of interest. The described procedure allows us to get the scattering coefficient of full field $ a_M^{(\mathrm{sc})}(0) = \tilde{a}_M^{(\mathrm{sc})}(0)$ and compute the magnetoelectric polarizability, which appears to be nontrivial for the radially magnetized cylinder:
\begin{equation} \label{eq:magnetoelectricpolariz}
    \begin{gathered}
         \alpha_{me}= \frac{1}{i \pi q^2 e_0} \frac{G'_0(ka)}{\sqrt{\varepsilon} {H'_0}^{(1)}(qa)} = \frac{1}{i \pi q^2 e_0} \frac{B J'_0(ka)+ \frac{g b^{(\text{cyl})}_E(0)}{\varepsilon}f'(ka)}{\sqrt{\varepsilon} {H'_0}^{(1)}(qa)} \\
    \end{gathered}
\end{equation}

Returning to the system \eqref{system2}, we analyze the second and third equations. Taking into account the equation for the radial function $P_0(x)$ [see third equation in \eqref{ODE2}], it is straightforward to obtain that $\tilde{a}_E^{(\mathrm{sc})}(0) = 0$ and $a_E^{(\mathrm{sc})}(0) = b_E^{(\mathrm{sc})}(0)$. This means that $\alpha_{ee}$ of a gyrotropic cylinder is exactly equal to $\alpha_{ee}$ of a dielectric cylinder, which is an expected result in the first order of the perturbation theory.

In a similar way we consider another polarization of the incident wave, when magnetic field is directed along the $z$ axis. In that case,
\begin{equation}
    \begin{gathered}
        \mathbf{H}^{(\mathrm{inc})} = h_0 \sum_{m=-\infty}^{+\infty} \left\{ i^mJ_m(q \rho) e^{i\,m \phi} \hat{\mathbf{z}} \right\} \, .\\
    \end{gathered}
\end{equation}
Using the perturbation theory for the fields inside and outside the cylinder, applying boundary conditions, considering multipole coefficients only with $m=0$ and accounting that $b_E^{(\mathrm{cyl})}(m) = b_E^{(\mathrm{sc})}(m) = 0$ for all $m$ (see Sec.~II), we derive the following system: ($\mu=1$):
\begin{equation}
    \begin{gathered} \label{systemTE}
        E_{\phi}: \quad \frac{1}{\sqrt{\varepsilon}}[b_M^{(\mathrm{cyl})}(0) J'_0(k a) + G'_0(k a)] = h_0 J'_0(q a) + [b_M^{(\mathrm{sc})}(0)+\tilde{a}_M^{(\mathrm{sc})}(0)] H_0'^{(1)}(q a) \,,\\
        E_{z}: \quad \frac{1}{\sqrt
        \eps}  P_0(ka) =   \tilde{a}_E^{(\mathrm{sc})}(0)H_0^{(1)}(q a) \,,\\
        H_{\phi}: \quad   P'_0(ka) =   \tilde{a}_E^{(\mathrm{sc})}(0)H_0'^{(1)}(q a)\,,\\
        H_{z}: \quad  b_M^{(\mathrm{cyl})}(0) J_0(k a) - \frac{G'_0}{ka}-G''_0(ka) = h_0 J_0(q a) + [b_M^{(\mathrm{sc})}(0) + \tilde{a}_M^{(\mathrm{sc})}(0)]H_0^{(1)}(q a).
    \end{gathered}
\end{equation}
The equations for the radial functions Eq.~\ref{ODE} are simplified in the case of TE illumination to the form:
\begin{equation} \label{ODE3}
    \begin{gathered}
         \left(\frac{F'_0}{x} + F''_0 + F_0\right)' = 0 \, ,\\
         \left(\frac{G'_0}{x}  + G''_0 + G_0\right)' = 0  \, ,\\ 
         \frac{P'_0}{x}  + P''_0 + P_0 = \frac{g b^{(\text{cyl})}_M(0)}{\varepsilon}J'_0(x) \, . 
    \end{gathered}
\end{equation}
From the first and fourth equations in the system \ref{systemTE}, taking into account the differential equation for the function $G_0(x)$ from \ref{ODE3}, we immediately obtain that $\tilde{a}_M^{(\mathrm{sc})}(0) = 0$ and $a_M^{(\mathrm{sc})}(0) = b_M^{(\mathrm{sc})}(0)$. Thus, in the first order of the perturbation theory,  $\alpha_{mm}$ of a gyrotropic cylinder is equal to $\alpha_{mm}$ of a dielectric cylinder.

To find the first correction to the scattering coefficient $a_E^{(\mathrm{sc})}(0)$, we need to solve the system
\begin{equation} \label{a_efinal}
 \begin{cases}
   \tilde{a}_E^{(\mathrm{sc})}(0) = \frac{P'_0(ka)}{\ H_0'^{(1)}(q a)},\\
   \frac{P'_0}{x}  + P''_0 + P_0 = \frac{g b^{(\text{cyl})}_M(0)}{\varepsilon}J'_0(x), \\
   \frac{1}{ \sqrt{\eps} }H_0'^{(1)}(q a)P_0(ka)-H_0^{(1)}(q a)P'_0(ka)=0.
 \end{cases}
\end{equation}
Since it is similar to the case of TM polarization discussed earlier, we will highlight only the most important points of this calculation. We are looking for the solution in the form of
\begin{equation} \label{formofP}
  P_0(x) = C\,J_0(x)+D+g(x),
\end{equation}
where $C\,J_0(x)$ is a homogeneous solution of the second equation in \eqref{a_efinal}, $g$ its inhomogeneous solution with the zero initial conditions and $D$ is an inessential additive constant. As before, the unknown $C$ constant is found from the boundary condition:
\begin{equation} \label{coeffC}
    \begin{gathered}
        C = - \frac{g b^{(\text{cyl})}_M(0)}{\varepsilon} \frac{g(ka) {H'_0}^{(1)}(qa)  - g'(ka) {H_0}^{(1)}(qa) \sqrt{\varepsilon}}{  J_0(ka) {H'_0}^{(1)}(qa)  - J'_0(ka) {H_0}^{(1)}(qa) \sqrt{\varepsilon}}.\\
    \end{gathered}
\end{equation}

Again, the numerical solution of the second equation in \eqref{a_efinal} could be problematic. To find the initial condition at some point close to zero, we decompose both parts of this equation into a series. In the leading order we get
\begin{equation} \label{functions}
    \begin{gathered}
        g \approx - \frac{1}{18} x^3 + \frac{17}{3600} x^5+...
    \end{gathered}
\end{equation}
This expansion allows us to set the condition at point $x=0$ and find the function $g(x)$ numerically. Finally, we compute $a_E^{(\mathrm{sc})}(0)=\tilde{a}_E^{(\mathrm{sc})}(0)$ and the desired electromagnetic polarizability:
\begin{equation} \label{eq:magnetoelectricpolariz}
    \begin{gathered}
        \alpha_{em}= \frac{1}{i \pi q^2 h_0} \frac{P'_0(ka)}{{H'_0}^{(1)}(qa)} = \frac{1}{i \pi q^2 h_0} \frac{ CJ'_0(ka)+\frac{g b^{(\text{cyl})}_M(0)}{\varepsilon} g'(ka)}{{H'_0}^{(1)}(qa)}.
    \end{gathered}
\end{equation}
Plotting the dependence of $\alpha_{em}$ and $\alpha_{me}$ on $w$ or $ka$, we notice that they match identically, which is a characteristic feature of the Tellegen response. This result is also consistent with the symmetry of the problem which prohibits chiral response of the cylinder.

\section{Checking the limiting case $ka\ll 1$}


In this section, we check that the expressions for the polarizabilities of the radially magnetized cylinder $\alpha_{em}=\alpha_{me}$, $\alpha_{mm}$  and $\alpha_{ee}$ derived in the main text match the approximate quasistatic expressions obtained above. 

The expressions for the polarizabilities of a gyrotropic cylinder provided in the main text have the form ($\mu = 1$):
\begin{equation} \label{eq:polarizabilities}
    \begin{gathered}
         \alpha_{ee}=  \frac{J_0(ka)J'_0(qa) - \sqrt{\varepsilon} J_0(qa) J'_0(ka)}{i \pi q^2( \sqrt{\varepsilon} {H_0}^{(1)}(qa)J'_0(ka) -  J_0(ka) {H'_0}^{(1)}(qa))},\\
         \alpha_{mm}= \frac{J_0(ka)J'_0(qa) \sqrt{\varepsilon} -  J_0(qa) J'_0(ka)}{i \pi q^2(  {H_0}^{(1)}(qa)J'_0(ka) -  \sqrt{\varepsilon}J_0(ka) {H'_0}^{(1)}(qa))},\\
         \alpha_{me}= \frac{1}{i \pi q^2 e_0} \frac{G'_0(ka)}{\sqrt{\varepsilon} {H'_0}^{(1)}(qa)} ,\\
        \alpha_{em}= \frac{1}{i \pi q^2 h_0} \frac{P'_0(ka)}{{H'_0}^{(1)}(qa)},
    \end{gathered}
\end{equation}
In the limit $ka\ll 1$ the following expansions are valid:
\begin{align} \label{asympt}
    &J_0(x) \approx 1 - \frac{x^2}{4},
    &H_0(x) \approx \frac{2 i}{\pi}\left[\ln(\frac{x}{2})+\gamma\right]+1\:,
\end{align}
where $\gamma$ is the Euler's constant. For electric and magnetic polarizabilities we obtain:
\begin{align} \label{1}
    & \alpha_{ee}  \approx  \frac{  1 \cdot (-\frac{qa}{2}) - \sqrt{\varepsilon}\cdot 1 \cdot (-\frac{ka}{2})}{ i \pi q^2 (\sqrt{\varepsilon} \cdot \frac{2i}{\pi}[\log(\frac{qa}{2})+\gamma] \cdot (-\frac{ka}{2}) -   1 \cdot \frac{2 i}{\pi q a})} \approx \frac{a^2(\varepsilon -1)}{4},  \\
    &\alpha_{mm}  \approx \frac{  1 \cdot (-\frac{qa}{2}) \cdot  \sqrt{\varepsilon} - 1 \cdot (-\frac{ka}{2}) }{i \pi q^2( \frac{2i}{\pi}[\log(\frac{qa}{2})+\gamma] \cdot (-\frac{ka}{2}) - \sqrt{\varepsilon} \cdot  1 \cdot \frac{2 i}{\pi q a}   )} \approx 0,
\end{align}
which is completely consistent with the quasistatic expressions.

The quasi-static limit of $\alpha_{em}$ and $\alpha_{me}$ is more sophisticated. For analysis, we write these expressions in more detail:
\begin{equation} \label{eq:magnetoelectricpolariz}
    \begin{gathered}
         \alpha_{me}= \frac{1}{i \pi q^2 e_0} \frac{G'_0(ka)}{\sqrt{\varepsilon} {H'_0}^{(1)}(qa)} = \frac{1}{i \pi q^2 e_0} \frac{B J'_0(ka)+ \frac{g b^{(\text{cyl})}_E(0)}{\varepsilon}f'(ka)}{\sqrt{\varepsilon} {H'_0}^{(1)}(qa)} ,\\
        \alpha_{em}= \frac{1}{i \pi q^2 h_0} \frac{P'_0(ka)}{{H'_0}^{(1)}(qa)} = \frac{1}{i \pi q^2 h_0} \frac{ CJ'_0(ka)+\frac{g b^{(\text{cyl})}_M(0)}{\varepsilon} g'(ka)}{{H'_0}^{(1)}(qa)},
    \end{gathered}
\end{equation}
where $B$ and $C$ coefficients have the form:
\begin{equation} \label{eq:polarizabilities}
    \begin{gathered}
        B= - \frac{g b^{(\text{cyl})}_E(0)}{\varepsilon}  \frac{ \sqrt{\varepsilon} {H'_0}^{(1)}(qa)  (\frac{f'(ka)}{ka}+f''(ka))+  {H_0}^{(1)}(qa) f'(ka)}{\sqrt{\varepsilon} {H'_0}^{(1)}(qa)  (\frac{J'_0(ka)}{ka}+J''_0(ka))+  {H_0}^{(1)}(qa) J'_0(ka)},\\
        C = - \frac{g b^{(\text{cyl})}_M(0)}{\varepsilon} \frac{g(ka) {H'_0}^{(1)}(qa)  - g'(ka) {H_0}^{(1)}(qa) \sqrt{\varepsilon}}{  J_0(ka) {H'_0}^{(1)}(qa)  - J'_0(ka) {H_0}^{(1)}(qa) \sqrt{\varepsilon}}.\\
    \end{gathered}
\end{equation}
Here, $b^{\text{cyl}}_E(0)$ and  $b^{\text{cyl}}_M(0)$ are the scattering coefficients  of a dielectric cylinder in the case of TM- and TE- polarization of the incident field respectively \ref{eq:scattcoef1}, \ref{eq:scattcoef2}. In the limit $ka \ll 1$, using \ref{asympt}, they take the form:
\begin{equation} \label{eq:scattcoef}
    \begin{gathered}
       b^{(\text{cyl})}_E(0) =  \frac{e_0 \sqrt{\varepsilon}(J_0(qa)H_0'^{(1)}(q a) -  J'_0(qa) H_0^{(1)}(q a))}{ J_0(ka) H_0'^{(1)}(q a) - \sqrt{\varepsilon} H_0^{(1)}(q a) J'_0(ka)}\approx \frac{e_0 \sqrt{\varepsilon}\frac{2i}{\pi qa}}{\frac{2i}{\pi qa}} \approx e_0 \sqrt{\varepsilon},\\
         b^{(\text{cyl})}_M(0) =  \frac{h_0 \sqrt{\varepsilon}(J_0(qa) H_0'^{(1)}(q a) - J'_0(qa)H_0^{(1)}(q a)) }{\sqrt{\varepsilon}  J_0(ka) H_0'^{(1)}(q a) - H_0^{(1)}(q a)J'_0(ka)} \approx \frac{h_0 \sqrt{\varepsilon}\frac{2i}{\pi qa}}{ \sqrt{\varepsilon}\frac{2i}{\pi qa}} \approx h_0,
    \end{gathered}
\end{equation}
Also the expressions \ref{eq:magnetoelectricpolariz}  contain the functions $f$ and $g$, which are inhomogeneous solutions of equations for the radial functions $G_0$ and $P_0$ with the zero initial conditions. The respective differential equations have the form:
\begin{equation} \label{eq:ODE}
    \begin{gathered}
         (\frac{G'_0}{x}  + G''_0 + G_0)' = \frac{g b^{\text{cyl}}_E(0)}{\varepsilon}J_0(x)  \, ,\\ 
         \frac{P'_0}{x}  + P''_0 + P_0 = \frac{g b^{\text{cyl}}_M(0)}{\varepsilon}J'_0(x) \, . 
    \end{gathered}
\end{equation}

In the exact solution of a ODE system,  functions $f$ and $g$ are found numerically, but for the analysis of the quasi-static case, the right-hand side of the differential equations can be expanded into a series and we can consider only the leading-order terms:
\begin{equation} \label{functions}
    \begin{gathered}
       f \approx \frac{x^3}{9}-\frac{7 x^5}{900}+...,  \ \ g \approx - \frac{1}{18} x^3 + \frac{17}{3600} x^5+...
    \end{gathered}
\end{equation}
Then, using \ref{eq:scattcoef} and \ref{functions}, we can calculate quasistatic expressions for B and C:
\begin{equation} \label{Constants}
    \begin{gathered}
        B \approx  - \frac{g e_0 \sqrt{\varepsilon}}{\varepsilon}  \frac{ \sqrt{\varepsilon} \frac{2 i}{\pi q a}  (\frac{ka}{3}+\frac{2 ka}{3})+  ...}{\sqrt{\varepsilon} \frac{2 i}{\pi q a}  (-\frac{1}{2}-\frac{1}{2})+  ...} \approx \frac{g e_0 ka}{\sqrt{\varepsilon}},\\
         C \approx - \frac{g h_0}{\varepsilon} \frac{-\frac{1}{18} (ka)^3 \frac{2 i}{\pi qa}+...}{  1 \cdot \frac{2i}{\pi qa}  + ...} \approx \frac{g h_0 (ka)^3}{18 \varepsilon }.\\
    \end{gathered}
\end{equation}
Substituting  \ref{eq:scattcoef}, \ref{Constants} into \ref{eq:magnetoelectricpolariz}, we recover the final expressions for $\alpha_{em}$ and $\alpha_{me}$ in the limit $ka \ll 1$:
\begin{equation} \label{eq:quasistatc}
    \begin{gathered}
         \alpha_{me}\approx   \frac{1}{i \pi q^2 e_0} \frac{\frac{g e_0 ka}{\sqrt{\varepsilon}} \frac{-ka}{2}+ \frac{g e_0}{\sqrt{\varepsilon}}  \frac{(ka)^2}{3}}{\sqrt{\varepsilon} \frac{2 i}{\pi q a}} \approx \frac{qga^3}{12} ,\\
        \alpha_{em} \approx  \frac{1}{i \pi q^2 h_0} \frac{ \frac{g h_0 (ka)^3}{18 \varepsilon } \frac{-ka}{2} +\frac{g h_0}{\varepsilon} \frac{-1 \cdot (ka)^2}{6}}{\frac{2 i}{\pi q a}} \approx \frac{qga^3}{12},
    \end{gathered}
\end{equation}
which is also consistent with the quasistatic expressions for magnetoelectric and electromagnetic polarizabilities derived in the section above.

\bibliography{main}